\documentclass[sort&compress,review]{elsarticle}
\usepackage{amsfonts,bbm,amssymb,amsmath,slashed,mathrsfs,graphicx,tikz,color,algorithm,algpseudocode}
\usepackage{lineno,hyperref}
\usepackage{lineno,hyperref,orcidlink}
\newtheorem{theorem}{Theorem}
\newtheorem{theoremsec}{Theorem}[section]
\newtheorem{lemma}{Lemma}[section]
\newtheorem{corollary}[theorem]{Corollary}

\newtheorem{example}{Example}
\newtheorem{question}{Open Problem}

\newdefinition{remark}{Remark}[section]
\newdefinition{definition}{Definition}[section]
\newdefinition{postulate}{Postulate}
\newproof{proof}{{\noindent\it Proof}}

\def\QEDopen{{\setlength{\fboxsep}{0pt}\setlength{\fboxrule}{0.2pt}\fbox{\rule[0pt]{0pt}{1.3ex}\rule[0pt]{1.3ex}{0pt}}}}
\def\QED{\QEDopen}

\def\Q.E.D{\hfill\QED}

\journal{}

\begin{document}

\begin{frontmatter}



\title{On Probabilistic $\omega$-Pushdown Systems, and $\omega$-Probabilistic Computational Tree Logic}

\author[a]{Deren Lin\corref{ir}}
\address[a]{Xiamen City, China}
\cortext[ir]{Independent researcher.}

\author[b]{Tianrong Lin\orcidlink{0000-0002-1187-2395}\corref{ca}}
\address[b]{Hakka University,  China}
\cortext[ca]{Corresponding author.}

\begin{abstract}
In this paper, we define the notion of a {\em probabilistic $\omega$-pushdown automaton} and study its model-checking problem against $\omega$-probabilistic computational tree logic ($\omega$-PCTL) and its bounded version from a computational complexity perspective. Specifically, we obtain the following  important new results:
\begin{itemize}
  \item [--]{We first discuss the expressiveness of the logics PCTL, PCTL$^*$, $\omega$-${\rm PCTL}$, and $\omega$-${\rm PCTL}^*$ and study how Büchi conditions of probabilistic $\omega$-pushdown systems influence $\omega$-PCTL formulas. We then investigate the model-checking problem for {\em stateless probabilistic $\omega$-pushdown system ($\omega$-pBPA)} against $\omega$-PCTL (as defined by Chatterjee, Sen, and Henzinger in \cite{CSH08}). By constructing $\omega$-PCTL formulas that encode the {\em Post Correspondence Problem}, we show that this model-checking problem is generally undecidable.}
      
  \item [--]{We then study under which conditions there exists an algorithm for model-checking {\it stateless probabilistic $\omega$-pushdown systems} against $\omega$-PCTL-like logic. In particular, we show that the model-checking problem for {\it stateless probabilistic $\omega$-pushdown systems} against $\omega$-{\it bounded probabilistic computational tree logic} ($\omega$-bPCTL) is decidable and $\mathit{NP}$-hard. Currently, there is no known lower bound for this problem that is better than ours.}

  \item [--]{Finally, we investigate an upper bound for the model-checking problem for {\em stateless probabilistic $\omega$-pushdown systems} against $\omega$-bounded probabilistic computational tree logic ($\omega$-bPCTL). We propose a potential approach to solving it by establishing a conditional upper bound and analyze the challenges of this method.}
\end{itemize}
\end{abstract}


\end{frontmatter}

\section{Introduction}
\label{sec:introduction}

As is well known, logic is the foundational and enduring topic of {\em theoretical computer science}. Dating back to 1936, one of Alan Turing's main goals in defining the Turing machine \cite{Tur37} was to address the Entscheidungsproblem (the decision problem). Today, logic continues to play a fundamental role in computer science. Some of the key areas of logic that are particularly significant include {\em computability theory}, {\em modal logic}, and {\em category theory}. More importantly, the {\em theory of computation} is largely built on concepts developed by logicians and mathematicians such as Alonzo Church \cite{Chu36a, Chu36b} and Alan Turing \cite{Tur37}, among others.

Over the last four decades, {\em Model checking} has become an essential tool for formal verification in the field of logic in computer science \cite{CGP99}. It plays a particularly important role in the verification of digital circuits (chips). In a typical model-checking task, one describes the system to be verified as a model in an appropriate logic, expresses the property to be verified as a formula in that logic, and then uses automated algorithms to check whether the formula holds in the model; see e.g., \cite{BK08,CGP99}. Specifically, it is an automatic method for verifying that a formal model of a system satisfies a given property. Traditionally, model checking has been applied to finite-state systems and non-probabilistic programs. As the technology has matured, researchers in computer science have increasingly focused over the last two decades on the model checking of probabilistic infinite-state systems; see e.g., \cite{EKM04, EKM06}.

To the best of our knowledge, one of the most notable examples of such probabilistic infinite-state systems is the probabilistic pushdown system, often referred to as probabilistic pushdown automata (with a singleton input alphabet) in the literature \cite{Bra07,BBFK14,EKM04,EKS03}. In this paper, we call this restricted model a probabilistic pushdown system (pPDS). In other words, probabilistic pushdown systems can be viewed as a limited version of the more general probabilistic pushdown automata, whose input alphabet may contain multiple symbols. The model-checking problem for these systems, first studied in \cite{EKM04}, has received significant attention; see, e.g., \cite{Bra07,BBFK14}. These works investigated the model checking of stateless probabilistic pushdown systems (pBPA) against PCTL$^*$ as well as the model checking of probabilistic pushdown systems (pPDS) against PCTL. However, despite these efforts, the problem of model-checking stateless probabilistic pushdown systems (pBPA) against PCTL has remained open (see e.g.,\cite{BBFK14}). To the best of our knowledge, this question was first posed in \cite{EKM04} (For the journal version, see \cite{EKM06}) and remained open until our recent work \cite{LL24}. In \cite{LL24}, we resolved this question by showing that model checking of stateless probabilistic pushdown systems against PCTL is undecidable.

We now shift our focus to {\em temporal logic}. As noted in \cite{EH86}, there are two main perspectives on the nature of time: the linear-time view, in which each moment has only one possible future, and the branching-time view, in which each moment may split into multiple possible futures. As the reader will see in the sequel, most results in this paper concern branching-time properties. However, the logic mentioned above, which is used to specify probabilistic branching-time properties, lacks the ability to express $\omega$-properties. A well-known extension of PCTL capable of expressing $\omega$-regular properties, called $\omega$-PCTL, was introduced by Chatterjee, Sen, and Henzinger in \cite{CSH08}. Later, Chatterjee, Chmel\'{i}k, and Tracol \cite{CCT16} studied partially observable Markov decision processes (POMDPs) under $\omega$-regular conditions specified via parity objectives. Indeed, the logic $\omega$-PCTL can express not only $\omega$-regular properties but also certain important probabilistic $\omega$-pushdown properties (though not all of them). Therefore, another key objective of this paper is to introduce the $\omega$-extension of probabilistic pushdown automaton, which we call probabilistic $\omega$-pushdown systems. Once this model is formally defined, we can investigate important problems such as model checking against $\omega$-PCTL. It is worth noting that there is another interesting $\omega$-extension of branching-time computational tree logic \cite{LL14}. However, extending that logic \cite{LL14} to the probabilistic setting appears difficult.

As mentioned earlier, the main objects of study in this paper are probabilistic $\omega$-pushdown systems and the logic $\omega$-PCTL (or its variants). The central goal of this work is to investigate the model-checking problem for probabilistic $\omega$-pushdown systems against several variants of $\omega$-PCTL from the perspective of computational complexity.

\subsection{Main Contributions}

Now let us introduce our new main contributions. We first clarify the expressiveness of the logics PCTL, PCTL$^*$, $\omega$-PCTL and $\omega$-PCTL$^*$ (see Theorem \ref{theoremfive}); then we extend the classical concept of {\em nondeterministic $\omega$-pushdown automata} to {\em probabilistic $\omega$-pushdown automata} and investigate how Büchi conditions influence $\omega$-PCTL formulas.At the same time, we show that there exist probabilistic $\omega$-pushdown specifications that cannot be expressed by $\omega$-PCTL.

Apart from that, in addition to discussing the expressiveness of the logics PCTL, PCTL$^*$, $\omega$-PCTL and $\omega$-PCTL$^*$ and how Büchi conditions influence $\omega$-PCTL formulas, we place our emphasis on the complexity of the model-checking problem for {\em stateless probabilistic $\omega$-pushdown systems} against $\omega$-PCTL and obtain the following important and interesting result:

\begin{theorem}
\label{theorem1}
Model checking of stateless probabilistic $\omega$-pushdown systems ($\omega$-pBPAs) against the logic $\omega$-PCTL is generally undecidable.
\end{theorem}

The following corollary is a clear and immediate consequence of Theorem \ref{theorem1}, since the logic $\omega$-PCTL is a sublogic of $\omega$-PCTL$^*$.

\begin{corollary}
\label{corollary2}
Model checking of stateless probabilistic $\omega$-pushdown systems ($\omega$-pBPAs) against the logic $\omega$-PCTL$^*$ is generally undecidable.
\end{corollary}

Further, the following corollary is stated (and proved) in Remark \ref{remark4.2} in Section \ref{sec:proof_of_theorem_1}:

\begin{corollary}
\label{corollary3}
Model checking of probabilistic $\omega$-pushdown systems ($\omega$-pPDSs) against the logic $\omega$-PCTL$^*$ is generally undecidable.
\end{corollary}

Next, we study to what extent the model-checking problem for stateless probabilistic $\omega$-pushdown systems against $\omega$-PCTL-like logics is decidable. Interestingly, we show that if we obtain the $\omega$-bounded probabilistic computational tree logic ($\omega$-bPCTL) by replacing the until operator (i.e., ${\bf U}$) in $\omega$-probabilistic computational tree logic ($\omega$-PCTL) with the bounded until operator (i.e., ${\bf U}^{\leq k}$), then the model-checking problem for stateless probabilistic $\omega$-pushdown systems against $\omega$-bPCTL is decidable. We further show that in this case the complexity of this problem is $\mathit{NP}$-hard. Thus, we have the following interesting result.

\begin{theorem}
\label{theorem2}
The model checking of stateless probabilistic $\omega$-pushdown systems ($\omega$-pBPA) against $\omega$-bounded probabilistic computational tree logic ($\omega$-bPCTL) is decidable, and it is $\mathit{NP}$-hard.
\end{theorem}

We would like to stress that there is currently no better lower bound for this problem than ours (see Remark \ref{remark5.2}).

We also study the problem of model checking stateless probabilistic $\omega$-pushdown systems against $\omega$-bounded probabilistic computational tree logic ($\omega$-bPCTL) and establish a conditional upper bound.

\begin{theorem}
\label{theorem.three}
For any $\omega$-bPCTL path formula $\phi$, if there exists a function $p(n)$ such that, for any $r=(Z)(\gamma_2)(\gamma_3)\cdots$ in $\triangle$, one can decide whether the prefix $(Z)(\gamma_2)\cdots (\gamma_{p(n)})$ of $r$ satisfies $$(Z)(\gamma_2)\cdots (\gamma_{p(n)})\models^{L}\phi,$$ where $n=|(\triangle,\phi)|$, then the model-checking problem for $\omega$-pBPA against $\omega$-bPCTL is in $\mathcal{DSPACE}(p(n))$.
\end{theorem}

Finally, we point out that proving a $\mathcal{PSPACE}$ upper bound for Theorem \ref{theorem.three} remains out of reach. The main obstacle is determining the exact value (or degree) of $p(n)$.

\subsection{Related Work}

During the last two decades, researchers in computer science have paid considerable attention to model checking of probabilistic infinite-state systems. The study of the model checking problem for probabilistic pushdown systems first appeared in \cite{EKM04}. We refer the reader to the references in \cite{EKM04} for old work on model checking probabilistic infinite-state systems. To the best of our knowledge, the article \cite{EKM04} is the first paper on model checking of probabilistic infinite-state systems (though we cannot rule out the possibility of earlier unpublished or overlooked work). Since \cite{EKM04,EKM06}, several papers have studied the model checking problem for probabilistic pushdown systems (pPDS) and stateless probabilistic pushdown systems (pBPA) against PCTL and PCTL$^*$, such as \cite{BBFK14}, which established the undecidability of model checking for pPDS against PCTL and for pBPA against PCTL$^*$. Recently, we answered in \cite{LL24} the long-standing open question of model checking stateless probabilistic pushdown systems against PCTL. This problem was first raised in \cite{EKM04} but remained open \cite{BBFK14} until our work \cite{LL24}.

The celebrated extension of PCTL capable of expressing $\omega$-regular properties, namely $\omega$-PCTL, was introduced by Chatterjee, Sen, and Henzinger in \cite{CSH08}. This logic is also central to the description of some properties of probabilistic $\omega$-pushdown systems in the present paper. The notions of probabilistic $\omega$-pushdown automaton and probabilistic $\omega$-pushdown system are introduced for the first time in this paper. Our extension builds upon the foundational work in \cite{CG77}.

\subsection{Organization}

The rest of this paper is structured as follows. In Section \ref{sec:preliminaries}, we review some basic definitions and fix the necessary notation. In Section \ref{sec:extension_pctl}, we first compare the expressiveness of the logics PCTL, PCTL$^*$, $\omega$-PCTL, and $\omega$-PCTL$^*$, and introduce probabilistic $\omega$-pushdown automata for the first time. In the same section, we also study how Büchi conditions influence $\omega$-PCTL formulas. The undecidability of model checking stateless probabilistic $\omega$-pushdown systems against $\omega$-PCTL was established in Section \ref{sec:proof_of_theorem_1}. In Section \ref{sec:proof_of_theorem2}, we prove the complexity results for model checking stateless probabilistic $\omega$-pushdown systems against $\omega$-bPCTL (i.e., Theorem \ref{theorem2}). In Section \ref{discussion_upper_bound}, we study the upper bound for the model checking problem of stateless probabilistic $\omega$-pushdown systems against $\omega$-bPCTL and prove Theorem \ref{theorem.three}. The final section presents the conclusions and discusses some directions for future research.

\section{Preliminaries}
\label{sec:preliminaries}

For the convenience of the reader, we make the paper self-contained. For elementary probability theory, the reader is referred to \cite{Shi95} by Shiryaev or \cite{Loe78a,Loe78b} by Lo\`{e}ve.

Let $\mathbb{N}_1=\{1,2,\cdots\}$ and $\mathbb{N}=\mathbb{N}_1\cup\{0\}$. For an $n\in\mathbb{N}_1$, $[n]$ denotes the set $\{1,\cdots, n\}$. Let $\mathbb{Q}$ be the set of all rational numbers. Let $|A|$ denote the cardinality of a finite set $A$. Let $\Sigma$ and $\Gamma$ denote non-empty finite alphabets. Then $\Sigma^*$ is the set of all finite words (including the empty word $\epsilon$) over $\Sigma$, and $\Sigma^+ = \Sigma^*\backslash \{\epsilon\}$. For any word $w\in \Sigma^*$, $|w|$ denotes its length, i.e., the number of symbols in it.

\subsection{Markov Chains}

Let us introduce the Markov chains first. Roughly, {\em Markov chains} are {\em probabilistic transition systems}, which are accepted as the most popular operational model for the evaluation of the performance and dependability of information-processing systems. For more details, see e.g., \cite{BK08}.

\begin{definition}[cf. \cite{BK08}]
\label{definition2.1}
A {\em (discrete) Markov chain} is a triple $\mathcal{M}=(S,{\bf P}, s_{init}, AP, L)$ where $S$ is a finite or countably infinite set of states, $${\bf P}: S\times S\rightarrow[0,1]$$ is a transition probability function such that for each $s\in S$:
$$
\sum_{t\in S}{\bf P}(s,t)=1.
$$
Here $s_{init}\in S$ is the initial state, $AP$ is a set of atomic propositions, and $L:S\rightarrow 2^{AP}$ a labeling function.
\end{definition}

Sometimes, in the sequel, we write $s\in\mathcal{M}$ to mean that $\mathcal{M} = (S, {\bf P}, s_{\rm init}, AP, L)$ is a Markov chain with $s \in S$.

$\mathcal{M}$ is called {\em finite} if $S$ and $AP$ are finite. A path in $\mathcal{M}$ is a finite or infinite sequence of states from $S$: $\pi = s_0 s_1 \cdots s_{n-1} \in S^n$ (or $s_0 s_1 \cdots \in S^{\omega}$) where $n \in \mathbb{N}_1$ such that ${\bf P}(s_i, s_{i+1}) > 0$ for each $i$. Let $Paths(\mathcal{M})$ denote the set of paths in $\mathcal{M}$, and $Paths_{fin}(\mathcal{M})$ (resp. $Paths_{infin}(\mathcal{M})$) denote the set of finite paths (resp. infinite paths) in $\mathcal{M}$. Similarly, $Paths_{fin}(s)$ (resp. $Paths_{infin}(s)$) denotes the set all finite paths (resp. infinite paths) starting in $s$.  

Except for $Paths_{infin}$, in what follows we often use $RUN$ to denote the set of all infinite paths in $\mathcal{M}$, and $RUN(\pi')$ to denote the set of all infinite paths starting with a given finite path $\pi'$. If an infinite path $\pi$ starts with a given finite path $\pi'$, we write $\pi' \in \mathsf{prefix}(\pi)$. Let $\pi$ be an infinite path. Then $\pi[i]$ denotes the $i$-th state $s_i$ of $\pi$, and $\pi_i$ denotes the suffix $s_i s_{i+1} \cdots$. Clearly, $\pi_1=\pi$ (We assume that $\pi$ begins at the state $\pi[1]$). Furthermore, a state $s'$ is {\em reachable} from a state $s$ if there exists a finite path starting in $s$ and ending at $s'$.

For each $s \in S$, $(RUN(s), \mathcal{F},\mathcal{P})$ is a probability space, where $\mathcal{F}$ is the $\sigma$-field generated by all basic cylinders $Cyl(\pi)$ for finite paths $\pi$ starting from $s$,
$$Cyl(\pi)=\{\widetilde{\pi}\in RUN(s) : \pi\in \mathsf{prefix}(\widetilde{\pi})\},$$ and $\mathcal{P} : \mathcal{F} \rightarrow [0, 1]$ is the unique probability measure such that $$\mathcal{P}(Cyl(\pi)) = \prod_{1\leq i\leq |\pi|-1} \mathcal{P}(s_i, s_{i+1})$$ where $\pi = s_1s_2\cdots s_{|\pi|}$ and $s_1=s$.

\subsection{Probabilistic Computational Tree Logic}
\label{subsec:pctl}

The logic PCTL was originally introduced in \cite{HJ94}, where the corresponding model-checking problem was focused mainly on {\em finite-state Markov chains}.

Let $AP$ be a fixed set of atomic propositions. Formally, the syntax of {\em probabilistic computational tree logic} PCTL is given by
$$\aligned
 \Phi&::={\bf true}\text{ $|$ }p\text{ $|$ }\neg\Phi\text{ $|$ }\Phi_1\wedge\Phi_2\text{ $|$ }\mathcal{P}_{\bowtie r}(\varphi)\\
 \varphi&::={\bf X}\Phi\text{ $|$ } \Phi_1{\bf U}\Phi_2
\endaligned$$

\noindent where $\Phi$ and $\varphi$ denote the state formulas and path formulas, respectively, and $p \in AP$ is an atomic proposition. In the above, $\bowtie$ ranges over the set
\begin{center}
$\{>,\geq, =,<,\leq\}$\footnote{ In fact, the comparison relations  ``$\geq$" and ``$=$" are sufficient enough for our discussion.},
\end{center}
and $r$ is a rational number with $0 \leq r \leq 1$.\footnote{The reason for $r$ being rational numbers is just algorithmic requirements.} 

Let $\mathcal{M} = (S,{\bf P}, s_{init}, AP, L)$ be a {\em Markov chain}, where $L: S \rightarrow 2^{AP}$ is an assignment (labeling function), and {\bf true} is the abbreviation of the always true. The semantics of PCTL over $\mathcal{M}$ is given by the following rules.
$$\aligned
\mathcal{M},s\models^L{\bf true}   \quad&\,\,\,\text{     }\quad\text{    for any $s\in S$}\\
\mathcal{M},s\models^L p                     \quad&{\rm iff}\quad\text{ $p\in L(s)$}\\
\mathcal{M},s\models^L\neg\Phi              \quad&{\rm iff}\quad\text{ $\mathcal{M},s\not\models^L\Phi$}\\
\mathcal{M},s\models^L\Phi_1\wedge \Phi_2   \quad&{\rm iff}\quad\text{ $\mathcal{M},s\models^L\Phi_1$ and $\mathcal{M},s\models^L\Phi_2$}\\
\mathcal{M},s\models^L\mathcal{P}_{\bowtie r}(\varphi)  \quad&{\rm iff}\quad\text{ $\mathcal{P}(\{\pi\in RUN(s):\mathcal{M},\pi\models^L\varphi\})\bowtie r$}\\
\mathcal{M},\pi\models^L{\bf X}\Phi \quad&{\rm iff}\quad\text{$\mathcal{M},\pi[1]\models^L\Phi$}\\
\mathcal{M},\pi\models^L\Phi_1{\bf U}\Phi_2 \quad&{\rm iff}\quad\text{$\exists k\geq 0$  such that $\mathcal{M},\pi[k]\models^L\Phi_2$ and $\forall j. 0\leq j<k:\mathcal{M},\pi[j]\models^L\Phi_1$}.
\endaligned$$

\begin{remark}
\label{remark2.3}
The logic PCTL$^*$ extends PCTL by removing the requirement that every temporal operator must be preceded by a state formula. Its path formulas are generated by the following syntax.
$$\aligned
\varphi&::=\Phi\text{ $|$ }\neg\varphi\text{ $|$ }\varphi_1\wedge\varphi_2\text{ $|$ }{\bf X}\varphi\text{ $|$ }\varphi_1{\bf U}\varphi_2.
\endaligned$$
\end{remark}

For the logic PCTL$^*$, we use the following abbreviations:
$$\aligned
{\bf F}a\overset{\rm def}{=}{\bf true}{\bf U}a\\
{\bf G}a\overset{\rm def}{=}\neg({\bf F}\neg a),
\endaligned$$ where $a\in AP$ is an atomic proposition.

The difference between PCTL and PCTL$^*$ is clear: every well-defined PCTL formula is also a well-defined PCTL$^*$ formula. However, the converse does not necessarily hold. The semantics of PCTL$^*$ path formulas over $\mathcal{M}$ are defined as follows:
$$\aligned
\mathcal{M},\pi\models^L\Phi\quad&{\rm iff}\quad\text{$\mathcal{M},\pi[0]\models^L\Phi$}\\
\mathcal{M},\pi\models^L\neg\varphi\quad&{\rm iff}\quad\text{$\mathcal{M},\pi\not\models^L\varphi$}\\
\mathcal{M},\pi\models^L\varphi_1\wedge\varphi_2\quad&{\rm iff}\quad\text{ $\mathcal{M},\pi\models^L\varphi_1$ and $\mathcal{M},\pi\models^L\varphi_2$}\\
\mathcal{M},\pi\models^L{\bf X}\varphi\,\quad&{\rm iff}\quad\mbox{$\mathcal{M},\pi_1\models^L\varphi$}\\
\mathcal{M},\pi\models^L\varphi_1{\bf U}\varphi_2\quad&{\rm iff}\quad\text{$\exists k\geq 0$ such that $\mathcal{M},\pi_k\models^L\varphi_2$ and $\forall j.0\leq j< k$: $\mathcal{M},\pi_j\models^L\varphi_1$}
\endaligned$$

\subsection{Post Correspondence Problem}
\label{sec:post_correspondence_problem}

The {\em Post Correspondence Problem} (PCP), originally introduced and shown to be undecidable by Post \cite{Pos46}, has been used to prove the undecidability of many problems in formal languages.

Formally, a PCP instance consists of a finite alphabet $\Sigma$ and a finite set $\{(u_i,v_i)\,:\,1\leq i\leq n\}\subseteq\Sigma^*\times\Sigma^*$ of $n$ pairs of strings over $\Sigma$. The problem asks whether there exists a non-empty sequence $j_1j_2\cdots j_k\in\{1,2,\cdots,n\}^+$ such that $u_{j_1}u_{j_2}\cdots u_{j_k}=v_{j_1}v_{j_2}\cdots v_{j_k}$.

There are numerous variants of the PCP. But the modified version from \cite{BBFK14} is the most convenient for our purposes. Since every word $w\in\Sigma^*$ has finite length, we may assume that $m=\max\{|u_i|,|v_i|\}_{1\leq i\leq n}$. Note that, to keep the problem undecidable, the number of pairs $n$, i.e., the number of tiles, must satisfy $\geq 5$ (see, e.g., \cite{A1,Nea15}).

If we insert the padding symbol ``$\bullet$" between letters (and possibly at the beginning or end) of each $u_i$ and $v_i$ to obtain padded strings $u'_i$ and $v'_i$ of equal length $m$, then the modified PCP asks whether there exists a sequence $j_1\cdots j_k\in \{1,\cdots,n\}^+$ such that $$u'_{j_1}\cdots u'_{j_k}=v'_{j_1}\cdots v'_{j_k}$$ after all ``$\bullet$" symbols are erased from the concatenated strings.

For example, consider the instance over $\Sigma'=\{A,B\}$ with $u_1=AB$, $v_1=BAB$, $u_2=AAB$, and $v_2=BB$. Then $$m=\max\{|u_i|,|v_i|\}_{i=1,2}=3.$$ A corresponding modified PCP instance over $\Sigma=\{A,B,\bullet\}$ is given by 
$$
u'_1=A\bullet B,\quad v'_1=BAB,\quad u'_2=AAB,\quad v_2'=\bullet BB.
$$
The question is whether there exists $j_1j_2\cdots j_k\in\{1,2\}^+$ such that $u'_{j_1}u'_{j_2}\cdots u_{j_k}'=v'_{j_1}v'_{j_2}\cdots v'_{j_k}$ after erasing all `$\bullet$' symbols.

\begin{remark}
\label{remark2.4}
Essentially, the modified PCP problem is equivalent to the original PCP problem. Padding the $n$ pairs of strings $u_i$ and $v_i$ with ``$\bullet$" symbols to make them the same length is useful in the following context for proving some of our main results.
\end{remark}

\subsection{Notions on Undecidability and $\mathit{NP}$-hard}

We introduce some notions in computational complexity used in this paper. For more information, we refer the reader to the excellent textbook \cite{DK14} or the beautiful lectures in computational complexity \cite{Cai03}.

Let $A$ be a decision problem. If there exists a Turing machine that decides $A$, then we say that the problem $A$ is decidable; otherwise, we say the problem $A$ is undecidable.

Let $\mathit{NP}$ denote the complexity class of all decision problems that are decidable in polynomial time by some nondeterministic Turing machine. A problem $A$ is $\mathit{NP}$-hard if every problem in $\mathit{NP}$ Karp-reduces to $A$. The problem $A$ is said to be $\mathit{NP}$-complete if $A\in NP$ and $A$ is $\mathit{NP}$-hard.

Other background information and notions will be introduced as needed when proving our main results stated in Section \ref{sec:introduction}.

\section{The $\omega$-PCTL and Probabilistic $\omega$-Pushdown Automata}
\label{sec:extension_pctl}

In this section, $\Sigma$ denotes a finite alphabet, and $\Sigma^*$ and $\Sigma^{\omega}$ denote the set of all finite words and the set of all $\omega$-sequences (or $\omega$-words) over $\Sigma$, respectively. An $\omega$-word over $\Sigma$ is written in the form $$\beta=\beta(1)\beta(2)\cdots$$ with $$\beta(i)\in\Sigma\text{ for all $i\in\mathbb{N}_1$}. $$ Let $\Sigma^{\infty}=\Sigma^*\cup\Sigma^{\omega}$. We use the following notation for segments of $\omega$-words: $$\beta(m,n):=\beta(m)\cdots \beta(n)\text{  (for $m\leq n$)};$$ and $$\beta(m,\omega):=\beta(m)\beta(m+1)\cdots. $$

For more details about $\omega$-words and $\omega$-languages, the reader is referred to the excellent works \cite{Sta97, Tho90}.

\subsection{$\omega$-PCTL}
\label{subsec:omega-pctl}

Now let us introduce the $\omega$-extension of PCTL defined in the celebrated work \cite{CSH08}. An obvious drawback of PCTL is that it cannot express important specifications such as liveness properties (i.e., the infinitely repeated occurrence of an event). The logic  $\omega$-PCTL can express such properties, and therefore has strictly greater expressiveness than PCTL. 

The formal syntax and semantics of $\omega$-PCTL are as follows. 

Let $AP$ be a fixed set of atomic propositions. The syntax of $\omega$-probabilistic computational tree logic ($\omega$-PCTL) is given by
$$\aligned
\Phi&::={\bf true}\text{ $|$ }p\text{ $|$ }\neg\Phi\text{ $|$ }\Phi_1\wedge\Phi_2\text{ $|$ }\mathcal{P}_{\bowtie r}(\varphi)\\
\varphi&::={\bf X}\Phi\text{ $|$ }\Phi_1{\bf U}\Phi_2\text{ $|$ }\varphi^{\omega}\\
\varphi^{\omega}&::=\text{Büchi}(\Phi)\text{ $|$ }\text{coBüchi}(\Phi)\text{ $|$ }\varphi_1^{\omega}\wedge\varphi_2^{\omega}\text{ $|$ }\varphi_1^{\omega}\vee\varphi_2^{\omega},
\endaligned$$
where $\Phi$ and $\varphi$ denote the state formulas and path formulas, respectively; $\varphi^{\omega}$ denotes infinitary path formulas (i.e., path formulas that depend on the set of states that appear infinitely often along a path); $p \in AP$ is an atomic proposition; $\bowtie\in\{>, \leq,>,\geq\}$; and $r$ is a rational number with $r\in\mathbb{Q}\cap[0,1]$.

The notion that a state $s$ (or a path $\pi$) satisfies a formula $\phi$ in a {\em Markov chain} $\widehat{M}$ is denoted by $\widehat{M},s\models^L\phi$ (or $\widehat{M},\pi\models^L\phi$) under a assignment (a labeling function) $L: S \rightarrow 2^{AP}$, and is defined inductively as follows:
$$\aligned
 \widehat{M},s\models^L {\bf true}  \quad&\text{  }\,\,\,\quad\text{    for any $s\in S$}\\
 \widehat{M},s\models^L p                     \quad&{\rm iff}\quad\mbox{ $p\in L(s)$}\\
 \widehat{M},s\models^L\neg\Phi              \quad&{\rm iff}\quad\mbox{ $\widehat{M},s\not\models^L\Phi$}\\
 \widehat{M},s\models^L\Phi_1\wedge \Phi_2   \quad&{\rm iff}\quad\mbox{ $\widehat{M},s\models^L\Phi_1$ and $\widehat{M},s\models^L\Phi_2$}\\
 \widehat{M},s\models^L\mathcal{P}_{\bowtie r}(\varphi)  \quad&{\rm iff}\quad\text{ $\mathcal{P}(\{\pi\in RUN(s):\widehat{M},\pi\models^L\varphi\})\bowtie r$}\\
 \widehat{M},\pi\models^L{\bf X}\Phi \quad&{\rm iff}\quad\text{$\widehat{M},\pi[1]\models^L\Phi$}\\
 \widehat{M},\pi\models^L\Phi_1{\bf U}\Phi_2 \quad&{\rm iff}\quad\text{$\exists k\geq 0$  such that $\widehat{M},\pi[k]\models^L\Phi_2$ and $\forall j. 0\leq j<k:\widehat{M},\pi[j]\models^L\Phi_1$}\\
  \widehat{M},\pi\models^L{\text{ Büchi}}(\Phi)\quad&{\rm iff}\quad\forall i\geq 0.\exists j\geq i. \text{ such that }\widehat{M},\pi[j]\models^L\Phi\\
  \widehat{M},\pi\models^L{\text{ coBüchi}}(\Phi)\quad&{\rm iff}\quad\exists i\geq 0.\forall j\geq i. \text{ such that }\widehat{M},\pi[j]\models^L\Phi\\
  \widehat{M},\pi\models^L\varphi_1^{\omega}\wedge\varphi_2^{\omega}\quad&{\rm iff}\quad \widehat{M},\pi\models^L\varphi_1^{\omega}\text{ and }\widehat{M},\pi\models^L\varphi_2^{\omega}\\
  \widehat{M},\pi\models^L\varphi_1^{\omega}\vee\varphi_2^{\omega}\quad&{\rm iff}\quad \widehat{M},\pi\models^L\varphi_1^{\omega}\text{ or }\widehat{M},\pi\models^L\varphi_2^{\omega}
\endaligned$$

\subsection{$\omega$-PCTL$^*$}

Similarly, one can define the logic $\omega$-PCTL$^*$. Note that $\omega$-PCTL$^*$ was not been defined in \cite{CSH08}.

 Formally, the syntax of $\omega$-probabilistic computational tree logic ($\omega$-PCTL$^*$) is given by
$$\aligned
\Phi&::={\bf true}\text{ $|$ }p\text{ $|$ }\neg\Phi\text{ $|$ }\Phi_1\wedge\Phi_2\text{ $|$ }\mathcal{P}_{\bowtie r}(\varphi)\\
\varphi&::=\Phi\text{ $|$ }\neg\varphi\text{ $|$ }\varphi_1\wedge\varphi_2\text{ $|$ }{\bf X}\varphi\text{ $|$ }\varphi_1{\bf U}\varphi_2\text{ $|$ }\varphi^{\omega}\\
\varphi^{\omega}&::=\text{Büchi}(\Phi)\text{ $|$ }\text{coBüchi}(\Phi)\text{ $|$ }\varphi_1^{\omega}\wedge\varphi_2^{\omega}\text{ $|$ }\varphi_1^{\omega}\vee\varphi_2^{\omega},
\endaligned$$
where $\Phi$ and $\varphi$ denote the state formulas and path formulas, respectively; $\varphi^{\omega}$ denotes infinitary path formulas(i.e., path formulas that depend on the set of states appearing infinitely often in a path); $p \in AP$ is an atomic proposition, $\bowtie\in\{>, \leq,>,\geq\}$; and $r$ is a rational number with $r\in\mathbb{Q}\cap[0,1]$.

The satisfaction relation for $\omega$-PCTL$^*$ formulas is defined similarly to the cases of $\omega$-PCTL and PCTL$^*$. That is, we write $\widehat{M},s\models^L\phi$ (resp. $\widehat{M},\pi\models^L\phi$) to mean that a state $s$ (resp. a path $\pi$) satisfies an $\omega$-PCTL$^*$ formula $\phi$ in the Markov chain $\widehat{M}$ under the labeling function $L:S\rightarrow 2^{AP}$. We omit others, but the semantics of $\omega$-PCTL$^*$ path formulas over $\widehat{M}$ are defined as follows:
$$\aligned
\widehat{M},\pi\models^L\Phi\quad&{\rm iff}\quad\text{$\widehat{M},\pi[0]\models^L\Phi$}\\
\widehat{M},\pi\models^L\neg\varphi\quad&{\rm iff}\quad\text{$\widehat{M},\pi\not\models^L\varphi$}\\
\widehat{M},\pi\models^L\varphi_1\wedge\varphi_2\quad&{\rm iff}\quad\text{ $\widehat{M},\pi\models^L\varphi_1$ and $\widehat{M},\pi\models^L\varphi_2$}\\
\widehat{M},\pi\models^L{\bf X}\varphi\,\quad&{\rm iff}\quad\mbox{$\widehat{M},\pi_1\models^L\varphi$}\\
\widehat{M},\pi\models^L\varphi_1{\bf U}\varphi_2\quad&{\rm iff}\quad\text{$\exists k\geq 0$ such that $\widehat{M},\pi_k\models^L\varphi_2$ and $\forall j.0\leq j< k$: $\widehat{M},\pi_j\models^L\varphi_1$}\\
  \widehat{M},\pi\models^L{\text{ Büchi}}(\Phi)\quad&{\rm iff}\quad\forall i\geq 0.\exists j\geq i. \text{ such that }\widehat{M},\pi[j]\models^L\Phi\\
  \widehat{M},\pi\models^L{\text{ coBüchi}}(\Phi)\quad&{\rm iff}\quad\exists i\geq 0.\forall j\geq i. \text{ such that }\widehat{M},\pi[j]\models^L\Phi\\
  \widehat{M},\pi\models^L\varphi_1^{\omega}\wedge\varphi_2^{\omega}\quad&{\rm iff}\quad \widehat{M},\pi\models^L\varphi_1^{\omega}\text{ and }\widehat{M},\pi\models^L\varphi_2^{\omega}\\
  \widehat{M},\pi\models^L\varphi_1^{\omega}\vee\varphi_2^{\omega}\quad&{\rm iff}\quad \widehat{M},\pi\models^L\varphi_1^{\omega}\text{ or }\widehat{M},\pi\models^L\varphi_2^{\omega}
\endaligned$$

\subsection{Expressiveness}

To describe the expressive power of the logics PCTL, PCTL$^*$, $\omega$-PCTL, and $\omega$-PCTL$^*$, we need the following definition.
\begin{definition}
Logic $L_2$ is at least as expressive as $L_1$ ($L_1 \preceq L_2$) if and only if 
$\forall \phi_1\in L_1$, there is a $\phi_2\in L_2$ such that for any Markov chain $M$ and a state $s\in M$ such that
$$
M,s\models\phi_1\Leftrightarrow M,s\models\phi_2.$$
\begin{enumerate}
  \item [(1)]{$L_2$ is strictly more expressive than $L_1$ ($L_1\prec L_2$) if $L_1\preceq L_2$ and there exists a formula $\phi_2\in L_2$ such that for no formula $\phi_1\in L_1$ and for any Markov chain $M$ and state $s\in M$, the following hold: 
      $$
      M,s\models\phi_2\Leftrightarrow M,s\models\phi_1.
      $$}
  \item [(2)]{Two logics are incomparable if neither is at least as expressive as the other (i.e., each can express some properties the other cannot).}
\end{enumerate}
\end{definition}

It is well-known that the PCTL$^*$ formula $\mathcal{P}_{\geq r}({\bf F}a\wedge{\bf G}b)$, where $r\in (0,1)$ and $a,b\in AP$, has no equivalent PCTL formula. Together with the fact that every PCTL formula is also a PCTL$^*$ formula, this implies that PCTL$^*$ is strictly more expressive than PCTL.

\begin{theorem}
\label{theoremfive}
The expressiveness relationships among PCTL, PCTL$^*$, $\omega$-PCTL, and $\omega$-PCTL$^*$ are as follows:
\begin{enumerate}
  \item [(1)]{$\omega$-PCTL $\succ$ PCTL;}
  \item [(2)]{$\omega$-PCTL$^*\succ\omega$-${\rm PCTL}$;}
  \item [(3)]{PCTL$^*\preceq\omega$-${\rm PCTL}^*$.}
\end{enumerate}
\end{theorem}
\begin{proof}
Item (1) follows directly from \cite{CSH08}. Since PCTL$^*\succ$ PCTL, there exists a formula $\phi\in\omega$-${\rm PCTL}^*$ (e.g., $\phi\overset{\rm def}{=}\mathcal{P}_{\geq r}({\bf F}a\wedge{\bf G}\mathcal{P}_{\leq r_1}(\text{Büchi}(b)))$, where $r,r_1\in(0,1)\cap\mathbb{Q}$ and $a,b\in AP$) that has no equivalent formula in $\omega$-${\rm PCTL}$. This proves Item (2). Item (3) is obvious and immediate.
\end{proof}

\begin{remark}
Since Büchi and coBüchi properties can be expressive in PCTL$^*$, we conjecture that PCTL$^*$ and $\omega$-${\rm PCTL}^*$ are equivalent expressive, with $\omega$-${\rm PCTL}^*$ simply providing a more direct and convenient syntax.
\end{remark}

\subsection{Probabilistic $\omega$-Pushdown Automata}

Let $\Gamma$ be a finite stack alphabet and $X\in\Gamma$. If $X\alpha\in\Gamma^+$, then the head of $X\alpha$, denoted by $head(X\alpha)$, is the symbol $X$. If $\gamma=\epsilon$, then $head(\gamma)=\epsilon$, where $\epsilon$ denotes the empty word. Furthermore, for a configuration $(q,X\gamma)$, the head is defined as $head(q,X\gamma)\overset{\rm def}{=}(q,head(X\gamma))$, i.e., $head(q,X\gamma)\overset{\rm def}{=}(q,X)$ (and $head(q,\epsilon)\overset{\rm def}{=}(q,\epsilon)$). 

There are several ways to define probabilistic $\omega$-pushdown automata. One approach starts with probabilistic pushdown automata and equips them with different acceptance conditions (see below). The other starts with nondeterministic $\omega$-pushdown automata and extends them with probabilities. Our method follows the second approach.

Before defining probabilistic $\omega$-pushdown automata, we first introduce (nondeterministic) $\omega$-pushdown automaton \cite{CG77}, on which our probabilistic model is based. The accepting conditions of our nondeterministic $\omega$-pushdown automata are similar to those in \cite{LSLZ17}.

\begin{definition}[cf. Definition 3.2.3 in \cite{CG77}]
\label{definition3.two}
A (nondeterministic) $\omega$-type pushdown automaton ($\omega$-PDA) is a $7$-tuple $M=(M_1,F)$, where $M_1=(Q,\Sigma,\Gamma,\delta,q_0,Z)$ is a nondeterministic pushdown machine with $Q$ a finite set of states, $\Sigma$ a finite input alphabet, $\Gamma$ a finite stack alphabet, $q_0\in Q$ the initial state, $Z\in\Gamma$ the initial stack symbol, and $\delta$ a transition function mapping $Q\times\Sigma\times\Gamma$ to finite subsets of $Q\times\Gamma^*$, and where $F$ is an acceptance condition defined below.
\end{definition} 

Let $w=\sigma_1\sigma_2\cdots\in\Sigma^{\omega}$ be an infinite word. We call an infinite sequence of configurations $r=\{(s_i,\gamma_i)\}_{i=1}^{\omega}$ a {\em run} of $M$ on $w$ if $(s_1,\gamma_1)=(q_0,Z)$ and $(s_i,\gamma_i)\overset{\sigma_i}{\Longrightarrow}(s_{i+1},\gamma_{i+1})$ for any $i\geq 1$. Every such run induces a mapping $$f_r:\mathbb{N}_1\rightarrow Q\times\Gamma, $$where $f_r(i)=(s_i,head(\gamma_i))$, i.e., the pair consisting of the state and head of the stack string $\gamma_i$ entering the $i$th step of the computation. For $(q,X)\in Q\times\Gamma$, we define the projection ${\rm Prj}_Q:Q\times\Gamma\rightarrow Q$ by $${\rm Prj}_{Q}(q,X)=q\in Q. $$ Now define ${\rm Inf}(r)$ to be the set of states that occur infinitely often in $r$, i.e., $${\rm Inf}(r)\overset{\rm def}{=}\{q:q={\rm Prj}_{Q}(f_r(i))\mbox{ for infinitely many }i\geq 1\}. $$ With the above notations, the acceptance condition $F$ can then be defined as follows:
\begin{enumerate}
  \item [(1)]{Büchi accepting condition (see e.g.,\cite{Buc66}): $F\subseteq Q$ is a set of states. A run $r$ is accepting iff ${\rm Inf}(r)\cap F\neq\emptyset$;}
  \item [(2)]{Rabin accepting condition (see e.g.,\cite{Rab69}): $F=\{(E_i,F_i)\}_{i=1}^k$ is a set of pairs with $E_i\subseteq Q$, $F_i\subseteq Q$. A run $r$ is accepting iff there exists $i$: $1\leq i\leq k$ such that $E_i\cap{\rm Inf}(r)=\emptyset$ and $F_i\cap{\rm Inf}(r)\neq\emptyset$;}
  \item [(3)]{Muller accepting condition (see e.g.,\cite{Mul63}): let $F=\{F_i\}_{i=1}^k$ be set of sets with $F_i\subseteq Q$. A run $r$ is accepting iff ${\rm Inf}(r)\in F$;}
  \item [(4)]{ Streett accepting condition (see e.g.,\cite{Str82}): $F=\{(E_i,F_i)\}_{i=1}^k$ is a set of pairs with $E_i\subseteq Q$, $F_i\subseteq Q$. A run $r$ is accepting iff for all $i$, $E_i\cap{\rm Inf}(r)\ne\emptyset$ or $F_i\cap{\rm Inf}(r)=\emptyset$.}
  \item [(5)]{Parity accepting condition (see e.g.,\cite{Mos85}): $k\in\mathbb{N}_1$ is a constant and $F:Q\rightarrow[k]$ is a priority function. A run $r$ is accepting iff $\min\{F(q)\,:\,q\in{\rm Inf}(r)\}$ is odd.}
\end{enumerate}

In this paper, we only consider the Büchi condition, which corresponds to Büchi formulas in the logic $\omega$-PCTL \cite{CSH08}.

By adding a probability distribution over the nondeterministic transitions in Definition \ref{definition3.two}, one obtains probabilistic transitions. This leads to the notion of a {\em probabilistic $\omega$-pushdown automaton}. 

Formally, a {\em probabilistic $\omega$-pushdown automaton} with Büchi accepting condition is defined as follows.

\begin{definition}
\label{definition3.1}
A {\em probabilistic $\omega$-pushdown automaton} is an $9$-tuple $\Theta=(Q,\Sigma,\Gamma,\delta,q_0,Z,F_{\rm inal},{\bf P}, \mathcal{F}) $ where
\begin{itemize}
\item {$Q$ is a finite set of states;}
\item {$\Sigma$ is a finite input alphabet;}
\item {$\Gamma$ is a finite stack alphabet;}
\item {$\delta$ is a mapping from $Q\times\Sigma\times\Gamma$ to finite subsets of $Q\times\Gamma^*$;}
\item {$q_0\in Q$ is the initial state;}
\item {$Z\in\Gamma$ is the initial stack symbol;}
\item {$F_{\rm inal}\subseteq Q$ is a set of (Büchi-accepting) states;}
\item {${\bf P}$ is a function that assigns to each transition rule $(p,a,X)\rightarrow(q,\alpha)\in\delta$ a probability
 $${\bf P}((p,a,X)\rightarrow(q,\alpha))\in[0,1]$$ such that for every $(p,a,X)\in Q\times\Sigma\times\Gamma$, $$\sum_{(q,\alpha)}{\bf P}((p,a,X)\rightarrow(q,\alpha))=1. $$ Without loss of generality, we assume $|\alpha|\leq 2$. The configurations of $\Theta$ are elements in $Q\times\Gamma^*$.}
\item {$\mathcal{F}$ is the Büchi accepting condition.}
\end{itemize}
\end{definition}

\begin{remark}
\label{remark3.1}
If there are multiple transition rules from the same configuration, we write them as
$$
(p,a,X)\rightarrow(q_1,\alpha_1)\,|\,(q_2,\alpha_2)\,|\,\cdots\,|\,(q_n,\alpha_n).
$$ 

The transition rule $(p,a,X)\rightarrow(q,\alpha)$ means that when the machine is in state $p$, the input symbol is $a$, and the top stack symbol is $X$, it moves to new state $q$ and replaces $X$ by the string $\alpha$ (see e.g., p. 228 of \cite{HMU07}). For example, if the current configuration is $(p,X\gamma)$ (with $X$ on top of the stack), applying the rule $(p,a,X)\rightarrow(q,\alpha)$ yields the new configuration $(q,\alpha\gamma)$.
\end{remark}

\begin{definition}
\label{definition3.2}
Let $\Theta=(Q,\Sigma,\Gamma,\delta,q_0,Z,F_{\rm inal},{\bf P},\mathcal{F})$ be a {\em probabilistic $\omega$-pushdown automaton}, and let $$\sigma=\prod_{i=1}^{\infty}a_i\in\Sigma^{\omega}$$ with $a_i\in\Sigma$, $\forall i\geq 1$. An infinite sequence of configurations $r=\{(s_i,\gamma_i)\}_{i\geq 1}$ is called a complete run of $\Theta$ on $\sigma$, starting in configuration $(s_1,\gamma_1)$, iff
\begin{enumerate}
\item {$(s_1,\gamma_1)=(q_0,Z)$;}
\item {for each $i\geq 1$, there exists $b_i\in\Sigma$ such that $$b_i:(q_i,\gamma_i)\rightarrow (q_{i+1},\gamma_{i+1})$$ and 
$$\prod_{i=1}^{\infty}b_i=\prod_{i=1}^{\infty}a_i.$$
}
\end{enumerate}
The run $r$ is called {\em successful} iff $\mathcal{F}$ is satisfied, i.e., $${\rm Inf}(r)\cap F_{\rm inal}\neq\emptyset. $$

Furthermore, we call an infinite sequence $$\pi=(q_0,Z)\overset{a_1}{\rightarrow}(s_2,\gamma_2)\overset{a_2}{\rightarrow}(s_3,\gamma_3)\overset{a_3}{\rightarrow}\cdots\in (Q\times\Gamma^*\times\Sigma)^{\omega}$$ a path of $\Theta$. We denote the projected $\omega$-word $a_1a_2\cdots\in\Sigma^{\omega}$ by ${\rm Prj}_{\Sigma}(\pi)$, i.e., $${\rm Prj}_{\Sigma}(\pi)\overset{\rm def}{=}a_1a_2\cdots\in\Sigma^{\omega}$$ and the sequence of configurations by $${\rm Prj}_{Q\times\Gamma^*}(\pi)\overset{\rm def}{=}(q_0,Z)(s_2,\gamma_2)(s_3,\gamma_2)\cdots \in (Q\times\Gamma^*)^{\omega}.$$

Let {\bf Path}$(q_0,Z)$ denote the set of all infinite paths of $\Theta$ with starting from configuration $(q_0,Z)$. And the $\omega$-word $\sigma\in\Sigma^{\omega}$ is said to be {\em accepted with probability at least $p$} (where $p\in[0,1]$) if $\mathcal{P}_{\Theta}(\sigma)\geq p$, where $$\mathcal{P}_{\Theta}(\sigma)=\mathcal{P}(\{\pi\in\mbox{\bf Path}(q_0,Z)\,:\,{\rm Prj}_{\Sigma}(\pi)=\sigma\,\bigwedge\,{\rm Inf}({\rm Prj}_{Q\times\Gamma^*}(\pi))\cap F_{\rm inal}\neq\emptyset\}). $$
\end{definition}

\begin{remark}
\label{remark3.2}
Given an input word $\sigma=a_1a_2\cdots\in\Sigma^{\omega}$, we define the scheduler $S(\sigma)$ such that $S(\sigma)((q_0,Z),\cdots,(s_{n-1},\gamma_{n-1}))(a_n)=1$. That is, at step $n$ the scheduler chooses with probability $1$ the letter $a_n$ as the next action. The operational behavior of $\Theta$ on reading the input word $\sigma$ is then captured by the Markov chain $\Theta_{S(\sigma)}$. We define the acceptance probability as
$$
\mathcal{P}_{\Theta}(\sigma)\overset{\rm def}{=}\mathcal{P}(\{\pi\in\mbox{\bf Path}(q_0,Z)\,:\,{\rm Prj}_{\Sigma}(\pi)=\sigma\,\bigwedge\,{\rm Inf}({\rm Prj}_{Q\times\Gamma^*}(\pi))\cap F_{\rm inal}\neq\emptyset\}).
$$
By \cite{CY95,Var85}, the set of accepting paths for word $\sigma$ is measurable.
\end{remark}

Now we define probabilistic $\omega$-pushdown systems. Intuitively, a probabilistic $\omega$-pushdown system is a degenerate form of a probabilistic $\omega$-pushdown automaton obtained by restricting the input alphabet to a singleton (a single input symbol, which is typically omitted). Its formal definition is as follows.

\begin{definition}
\label{definition3.3}
  A {\em probabilistic $\omega$-pushdown system ($\omega$-pPDS)} is a $8$-tuple $\Theta'=(Q,\Gamma,\delta,q_0,Z, F_{\rm inal},{\bf P},\mathcal{F}) $, whose configurations are elements $\in Q\times\Gamma^*$, where $\Gamma$ is a finite stack alphabet, $\delta$ a finite set of rules fulfilling
  \begin{itemize}
\item {$Q$ is a finite set of states;}
\item {$\Gamma$ is a finite stack alphabet and $\delta$ is a finite set of rules such that for each $(p,X)\in Q\times\Gamma$, there is at least one rule of the form $((p,X),(q,\alpha))\in\delta$ with $\alpha\in\Gamma^*$. We write $(p,X)\rightarrow(q,\alpha)$ instead of $((p,X),(q,\alpha))\in\delta$. We assume without loss of generality that $|\alpha|\leq 2$.}
\item {${\bf P}$ is a probability function that assigns to each rule $(p,X)\rightarrow(q,\alpha)$ in $\delta$ a value $${\bf P}((p,X)\rightarrow(q,\alpha))\in[0,1] $$ such that for every $(p,X)\in Q\times\Gamma$,
     $$\sum_{(q,\alpha)}{\bf P}((p,X)\rightarrow(q,\alpha))=1.$$
}
\item{ $F_{\rm inal}\subseteq Q$ is a set of (Büchi-accepting) states;}
\item{$\mathcal{F}$ is the Büchi accepting condition.}
\end{itemize}
\end{definition}

Let $\Theta'$ be a probabilistic $\omega$-pushdown system. An infinite sequence of configurations $r=\{(s_i,\gamma_i)\}_{i\geq 1}$ is called a complete run of $\Theta'$, starting from configuration $(s_1,\gamma_1)$, iff
\begin{enumerate}
    \item {$(s_1,\gamma_1)=(q_0,Z)$;}
    \item {for each $i\geq 1$, $(s_i,\gamma_i)\rightarrow (s_{i+1},\gamma_{i+1})$.
}
\end{enumerate}
Every such run induces a mapping $f_r:\mathbb{N}_1\rightarrow Q\times\Gamma$, where $$f_r(i)=(s_i,{\rm head}(\gamma_i)), $$entered in the $i$th step of the computation described by run $r$. The run $r$ is called {\em successful} if $${\rm Inf}(r)\cap F_{\rm inal}\neq\emptyset. $$

Furthermore, we call an infinite sequence $$\pi=(q_0,Z)(s_2,\gamma_2)\cdots\in (Q\times\Gamma^*)^{\omega}$$ a path. Let {\bf Path}$(q_0,Z)$ denote the set of all infinite paths of $\Theta'$ with starting from configuration $(q_0,Z)$.

The notion of a {\em stateless probabilistic $\omega$-pushdown systems} is obtained naturally from the definition above by restricting the state set $Q$ to a singleton (which is typically omitted). 

However, when $Q$ is a singleton the standard Büchi accepting condition becomes trivial, since the single state occurs infinitely often in every infinite run. To obtain a meaningful model, we therefore adjust the acceptance condition to depend on the heads of the stack configurations. Concretely, we take $F_{\rm inal}\subseteq\Gamma\cup\{\epsilon\}$. 

We are now ready to definite stateless probabilistic $\omega$-pushdown systems ($\omega$-pBPA).

\begin{definition}
\label{definition3.4}
A {\em stateless probabilistic $\omega$-pushdown system ($\omega$-pBPA)} is a $6$-tuple $\Theta'= (\Gamma, \delta,Z,F_{\rm inal},{\bf P},\mathcal{F})$, whose configurations are elements $\in\Gamma^*$, where $\Gamma$ is a finite stack alphabet, $\delta$ a finite set of rules satisfying
\begin{itemize}
\item {$\Gamma$ is a finite stack alphabet and $\delta$ is a finite set of rules such that for each $X\in\Gamma$ there is at least one rule $(X,\alpha)\in\delta$ with $\alpha\in\Gamma^*$. We write $X\rightarrow\alpha$ instead of $(X,\alpha)\in\delta$. We assume without loss of generality that $|\alpha|\leq 2$.}
\item {${\bf P}$ is a probability function that assigns to each rule $X\rightarrow\alpha$
in $\delta$ a value ${\bf P}(X\rightarrow\alpha)\in[0,1]$ such that for every $X\in\Gamma$ $$\sum_{\alpha}{\bf P}(X\rightarrow\alpha)=1.$$
}
\item{ $F_{\rm inal}\subseteq \Gamma\cup\{\epsilon\}$ is the set of (Büchi-accepting) heads;}
\item{ $\mathcal{F}$ is the Büchi accepting condition.}
\end{itemize}
\end{definition}

Since the state is fixed in a stateless system, we omit it. Thus, a run can be written as an infinite sequence of $r=\{(\gamma_i)\}_{i \geq 1}$ of stack contents. Similarly, every such infinite sequence $r=\{(\gamma_i)\}_{i \geq 1}$ induces a mapping $f_r:\mathbb{N}_1\rightarrow\Gamma\cup\{\epsilon\}$, where $f_r(i)={\rm head}(\gamma_i)$, i.e., the head symbol of the stack configuration $\gamma_i$ at the $i$-th step of the computation. Because we have modified the acceptance condition in our Definition \ref{definition3.4}, a run $r$ is now called {\em successful} if 
$${\rm Inf}(r)\cap F_{\rm inal}\neq\emptyset,$$where
$${\rm Inf}(r)=\{\gamma\in\Gamma\cup\{\epsilon\}\,:\,\gamma=f_r(i)\text{ for infinitely many $i\geq 1$}\}.$$

A complete run of $\Theta'$ starting from configuration $(\gamma_1)$ satisfies $\gamma_1=Z$ and $\gamma_i\rightarrow\gamma_{i+1}$ for all $i\geq 1$. We call an infinite sequence $\pi=(Z)(\gamma_i)\cdots\in(\Gamma^*)^{\omega}$ a path. Let {\bf Path}$(Z)$ denote the set of all infinite paths of $\Theta'$ starting from configuration $(Z)$.

As shown in \cite{EKS03} for the probabilistic setting, undecidable properties can be easily encoded into pushdown configurations without effective valuation assumptions. Therefore, throughout this paper we adopt the simple assignment used in \cite{EKS03,EKM06,BBFK14}, defined as follows.

\begin{definition}[simple assignment]\footnote{Our definition is essentially consistent with Definition 2.3 in \cite{BBFK14}.}
\label{definition3.5}
The head of a configuration $(p,\gamma)\in Q\times\Gamma^*$, denoted by $head(p,\gamma)$, is $(p,X)$ if $\gamma=X\alpha$ for some $X\in\Gamma$ and $\alpha\in\Gamma^*$, and $p$ if $\gamma=\epsilon$. For $\gamma=X\alpha\in\Gamma^*$, we also write $head(\gamma)=X$. A labeling function $L: Q\times\Gamma^*\rightarrow 2^{AP}$ is called a {\em simple assignment} if for each $a\in AP$ there exist a set of heads $H_a\subseteq Q\cup (Q\times\Gamma)$ such that with $A\subseteq AP$
$$L((p,X\alpha)=A  \text{ iff } head(p,X\alpha)\in\bigcup_{a\in A}H_a.$$
\end{definition}

Given an $\omega$-pPDS or $\omega$-pBPA $\triangle$, its configurations and transition rules induce an {\em infinite-state Markov chain} $\widehat{M}_{\triangle}$. The model-checking problem for an $\omega$-PCTL formula $\Psi$ (that is not a PCTL formula) is to decide whether $$\widehat{M}_{\triangle},Z\models^L\Psi, $$ where $L$ is a simple assignment (see Definition \ref{definition3.5}).

\subsection{How Büchi Conditions of Probabilistic $\omega$-Pushdown Systems Influencing $\omega$-PCTL Formulas}
\label{subsection.three.five}

In this subsection, we study how Büchi acceptance condition of a probabilistic $\omega$-pushdown system influence a given $\omega$-PCTL formula. 

First, we show that there exist probabilistic $\omega$-pushdown properties that cannot be expressed by any $\omega$-PCTL formula.

\begin{example}
\label{example1}
Let $AP\overset{\rm def}{=}\{cal,ret\}$. let $\#a$ denote the number of occurrences of the atomic proposition $a\in AP$. Let the path formula $\phi$ be defined as
$$\aligned
\phi\overset{\rm def}{=}&\text{``the trace is perfectly balanced (never more cal than ret in any prefix)}\\
&\text{ and the stack returns to empty infinitely often"}.
\endaligned$$
The formula $\phi$ is an infinite-path property that extends the one from \cite{A2}. That is, on an infinite path $\pi=s_1s_2\cdots$, the following hold: 
\begin{enumerate}
  \item [(a)]{ for every finite prefix $s_1\cdots s_n$, $\#cal\geq \#ret$ (the running stack depth never goes negative);}
  \item [(b)]{ there are infinitely many positions i where the stack depth is exactly 0 after $s_i$, i.e., $\#cal=\#ret$ in the prefix $s_1\cdots s_i$.}
\end{enumerate}
\end{example}

It is easy to show that $\phi$ in Example \ref{example1} cannot be expressed by any path formula of $\omega$-PCTL. Consequently, $\mathcal{P}_{\geq 1}(\phi)$ is not a state formula of $\omega$-PCTL. Hence no $\omega$-PCTL formula defines exactly the models of $\mathcal{P}_{\geq 1}(\phi)$. The above arguments immediately yield the following result.

\begin{theorem}
\label{theorem.seven}
The class of probabilistic $\omega$-pushdown specifications is a strict superset of the class of specifications definable by $\omega$-PCTL formulas.
\end{theorem}

Next, we study how the Büchi accepting condition of a probabilistic $\omega$-pushdown system influences whether it satisfies a given $\omega$-PCTL formula.

The Büchi accepting condition in a probabilistic $\omega$-pushdown system acts as a powerful filter that determines the set of infinite runs considered when evaluating an $\omega$-PCTL formula. Specifically, the Büchi accepting condition selects a subset of possible infinite runs. Only these accepting runs contribute to the probability measure when checking $\omega$-PCTL formulas. $\omega$-PCTL extends PCTL with infinitary path operators such as ${\rm B\text{ü}chi}(\phi)$, ${\rm coB\text{ü}chi}(\phi)$, etc., which allow it to express properties about sets of states (or stack heads) visited infinitely often. The influence of the Büchi accepting condition is fundamental:

\begin{enumerate}
  \item {({\bf Path filtering:}) The probability in $\mathcal{P}_{\geq p}(\Psi)$ (where $\Psi$ is a path formula of $\omega$-PCTL) is computed only over the accepting runs (those satisfying the Büchi accepting condition). Non-accepting runs are discarded (i.e., probability mass outside the accepted language is ignored or treated as non-satisfying, depending on the exact semantics).}
  \item {({\bf Interaction with stack:}) The stack enforces context-free aspects (e.g., proper nesting of calls/returns in Example \ref{example1}), while the Büchi accepting condition (on control states or stack heads) enforces $\omega$-regular properties.}
  \item {({\bf Probability measure:}) The overall probability that the system satisfies a given path formula is the measure of those accepting runs (under the probabilistic transitions) that also satisfy the temporal requirement expressed by the $\omega$-PCTL formula.}
\end{enumerate}

In short, the Büchi accepting condition defines the model (the set of valid infinite behaviors) in probabilistic $\omega$-pushdown systems. It directly controls which runs contribute to the probability calculation in $\omega$-PCTL, making it the key mechanism that links context-free stack behavior with $\omega$-regular properties. This combination yields a highly expressive formalism.

\section{Undecidability of Model-Checking of $\omega$-pBPA against $\omega$-PCTL}
\label{sec:proof_of_theorem_1}

Our goal in this section is to establish a theorem concerning the model-checking problem for stateless probabilistic $\omega$-pushdown systems against $\omega$-PCTL, which is conjectured to be undecidable. Clearly, the most straightforward way to prove undecidability is to show that we can encode a modified {\em Post Correspondence Problem} into a path formula of $\omega$-PCTL.

To this end, let $\Sigma = \{A, B, \bullet\}$ and the stack alphabet $\Gamma$ of the {\em $\omega$-pBPA} be
$$\Gamma=\{Z,Z',C,F,S,N\}\bigcup\left(\Sigma\times\Sigma\right)\bigcup\{X_{(x,y)}\,:\,(x,y)\in\Sigma\times\Sigma\}\bigcup\{G_i^j\,:\,,1\leq i\leq n,1\leq j\leq m+1\} $$ 

The elements of $\Gamma$ also serve as atomic propositions. We will describe how to construct the desired {\em $\omega$-pBPA} $\Theta'=(\Gamma, \delta,Z,F_{\rm inal}=\{Z'\},{\bf P},\mathcal{F})$.

Similar to the construction in \cite{LL24} (or similar to \cite{BBFK14}; see the references therein for more details), our {\it $\omega$-pBPA} $\Theta'$ operates in two phases, the first phase guesses a possible solution to a modified PCP instance by pushing pairs of words $(u_i,v_i)$ onto the stack, according to the following transition rules (with uniformly distributed probabilities):
\begin{equation}
\label{eq1}
\begin{split}
Z&\rightarrow G_1^1Z'\,|\,\cdots\,|\,G_n^1Z';\\
G_i^j&\rightarrow G_i^{j+1}(u_i(j),v_i(j));\\
G_i^{m+1}&\rightarrow C\,|\,G_1^1\,|\,\cdots \,|\,G_n^1.
\end{split}
\end{equation}

Concretely, according to (\ref{eq1}), $Z$ serves as the initial stack symbol. It begins with pushing $G_i^1Z'$ ($\in\Gamma^*$ for $1\leq i\leq n$) into the stack with probability $\frac{1}{n}$. The symbol $G_i^1$ (we read the stack from left to right) is then successively replaced (with probability $1$) by $G_i^2(u_i(1),v_i(1))$, and so on, until $G_i^{m+1}(u_i(m),v_i(m))$ \footnote{Note again that the modified PCP instance over $\Sigma$ has the property that $|u_i|=|v_i|=m$ for all $1\leq i\leq n$.} appears on top of the stack. This completes the storage of the first pair $(u_i,v_i)$.

At this point, with probability $\frac{1}{n+1}$, the system either pushes $C$ (to terminate the guessing phase) or pushes another $G_i^1$ (to continue guessing). When the rule $G_i^{m+1}\rightarrow C$ is applied, $\Theta'$ proceeds to the verification phase to check whether the sequence of pairs stored on the stack forms a solution to the modified PCP instance.

The guessing phase produces a word $j_1j_2\cdots j_k\in\{1,\cdots,n\}^+$ that corresponds to the sequence of pairs $(u_{j_1},v_{j_1}),\cdots,(u_{j_k},v_{j_k})$ pushed onto the stack (in that order). No other transition rules except those illustrated by (\ref{eq1}) are available during the guessing phase. This construction yields the following lemma:

\begin{lemma}[cf. \cite{BBFK14}; also \cite{LL24}]
\label{lemma4.1}
A configuration of the form $C\alpha Z'$ of $\Theta'$ is reachable from the initial configuration $Z$ if and only if $\alpha\equiv(x_1,y_1)\cdots(x_l,y_l)$ where $x_j,y_j\in\Sigma$ for all $j$, and there is a word $j_1j_2\cdots j_k\in\{1,2,\cdots,n\}^+$ such that $x_l\cdots x_1=u_{j_1}\cdots u_{j_k}$ and $y_l\cdots y_1 = v_{j_1}\cdots v_{j_k}$. Moreover, the probability of reaching $C\alpha Z'$ from $Z$ is $>0$. \Q.E.D
\end{lemma}

The next step is for $\Theta'$ to verify whether the stored sequence of pairs forms a solution. The corresponding transition rules (with uniformly distributed probabilities) are as follows:
\begin{equation}
\label{eq2}
\begin{split}
C&\rightarrow N\\
N&\rightarrow F\,|\,S\\
F&\rightarrow\epsilon\\
S&\rightarrow\epsilon\\
(x,y)&\rightarrow X_{(x,y)}\,|\,\epsilon\\
X_{(x,y)}&\rightarrow\epsilon\\
Z'&\rightarrow Z'
\end{split}
\end{equation}
This phase differs slightly from the one given in \cite{LL24}. Specifically, we replace the original rule $$Z'\rightarrow X_{(A,B)}\,|\,X_{(B,A)}$$ with the rule $$Z'\rightarrow Z'$$ in order to construct a rigorous $\omega$-PCTL state formula. Let $r=(Z)(\gamma_2)(\gamma_3)\cdots$ be a run of $\Theta'$. The Büchi accepting condition requires ${\rm Inf}(r)\cap F_{\rm inal}\neq\emptyset$, i.e., that $r$ is successful.

We define the following two $\omega$-PCTL path formulas: 
\begin{equation}
\label{eq3}
\begin{split}
\Psi_1 &= (\neg S\wedge\bigwedge_{z\in\Sigma}\neg X_{(B,z)}){\bf U}\left(\Big[\bigvee_{z\in\Sigma}X_{(A,z)}\Big]\vee \Big[Z'\wedge\mathcal{P}_{=1}\Big({\rm B\text{ü}chi}(Z')\Big)\Big]\right)\\
\Psi_2 &=(\neg F\wedge\bigwedge_{z\in\Sigma}\neg X_{(z,A)}){\bf U}\left(\Big[\bigvee_{z\in\Sigma}X_{(z,B)}\Big]\vee \Big[Z'\wedge\mathcal{P}_{=1}\Big({\rm B\text{ü}chi}(Z')\Big)\Big]\right)
\end{split}
\end{equation}
Note that the sub-formula $Z'\wedge\mathcal{P}_{=1}({\rm B\text{ü}chi}(Z')$ is a valid $\omega$-PCTL state formula. Moreover, since $Z'\rightarrow Z'$ holds with probability $1$, we have $$Z'\models^L\mathcal{P}_{=1}\Big({\rm B\text{ü}chi}(Z')\Big). $$

We now  prove Theorem \ref{theorem1}. Similar to \cite{LL24}, we define functions $\vartheta$, $\overline{\vartheta}$, $\rho$, and $\overline{\rho}$ and establish the following auxiliary result:
\begin{lemma}
\label{lemma4.2}
Let $\vartheta$ and $\overline{\vartheta}$ be two functions from $\{A,B,Z'\}$ to $\{0,1\}$, defined by
$$
             \vartheta(x)=\left\{
                              \begin{array}{ll}
                                1, & \hbox{$X=Z'$;} \\
                                1, & \hbox{$X=A$;} \\
                                0, & \hbox{$X=B$.}
                              \end{array}
                            \right.\,\,\,
             \overline{\vartheta}(x)=\left\{
                                         \begin{array}{ll}
                                           1, & \hbox{$X=Z'$;} \\
                                           0, & \hbox{$X=A$;} \\
                                           1, & \hbox{$X=B$.}
                                         \end{array}
                                       \right.
$$

Further, let $\rho$ and $\overline{\rho}$ be functions from $\{A,B\}^+Z'$ to $[0,1]$, given by
$$
 \rho(x_1x_2\cdots x_n)\overset{\rm def}{=}\sum_{i=1}^n\vartheta(x_i)\frac{1}{2^i},\quad
 \overline{\rho}(x_1x_2\cdots x_n)\overset{\rm def}{=}\sum_{i=1}^n\overline{\vartheta}(x_i)\frac{1}{2^i}.
$$

Then, for any $(u'_{j_1},v'_{j_1}),(u'_{j_2},v'_{j_2}),\cdots,(u'_{j_k},v'_{j_k})\in\{A,B\}^+\times\{A,B\}^+$,
$$
u'_{j_1}u'_{j_2}\cdots u'_{j_k} = v'_{j_1}v'_{j_2}\cdots v'_{j_k}
$$
if and only if
$$
\rho(u'_{j_1}\cdots u'_{j_k}Z')+\overline{\rho}(v'_{j_1}v'_{j_2}\cdots v'_{j_k}Z') = 1
$$
\end{lemma}
\begin{proof}
The proof is similar to that in \cite{LL24} and is therefore omitted.
\end{proof}

Let ${\rm trim}(b_1b_2\cdots b_n)$ denote the word $\in\{A,B\}^*$ obtained by removing all occurrences of `$\bullet$' from $b_1b_2\cdots b_n$. We now prove the following lemma:

\begin{lemma}
\label{lemma4.3}
Let $\alpha = (u_{j_1},v_{j_1})(u_{j_2},v_{j_2})\cdots (u_{j_k},v_{j_k})$ be the sequence of pairs pushed onto the stack by $\Theta'$, and let $(u'_{j_i},v'_{j_i})$, $1\leq i\leq k$, be the pair obtained from $(u_{j_i},v_{j_i})$ by erasing all ``$\bullet$" symbols, for $1\leq i\leq k$. Then
$$\aligned
\mathcal{P}(\{\pi\in\text{$RUN(F\alpha Z')$}\,:\,\pi\models^L\Psi_1\})=&\rho(u'_{j_1}u'_{j_2}\cdots u'_{j_k}Z')\\
            \mathcal{P}(\{\pi\in\text{$RUN(S\alpha Z')$}\,:\,\pi\models^L\Psi_2\})=&\overline{\rho}(v'_{j_1}v'_{j_2}\cdots v'_{j_k}Z').
\endaligned$$
\end{lemma}

\begin{proof}
Let $\mathcal{P}(F\alpha Z',\Psi_1)$ and $\mathcal{P}(S\alpha Z',\Psi_2)$ be shorthand notations for the probabilities above. Suppose $$u_{j_1}u_{j_2}\cdots u_{j_k}=x_1x_2\cdots x_l$$ and $$v_{j_1}v_{j_2}\cdots v_{j_k}=y_1y_2\cdots y_l. $$ We prove by induction on $l$ (i.e., the length of $\alpha$) that $$\mathcal{P}(F\alpha Z',\Psi_1)=\rho({\rm trim}(x_1x_2\cdots x_l)Z'), $$with a symmetric argument for $\Psi_2$.

By the rules in (\ref{eq2}), $F\alpha Z'\rightarrow\alpha Z'$ with probability $1$, so $\mathcal{P}(F\alpha Z',\Psi_1)=\mathcal{P}(\alpha Z',\Psi_1)$. Thus, it suffices to show $\mathcal{P}(\alpha Z',\Psi_1)=\rho({\rm trim}(x_1\cdots x_l)Z')$. 
We proceed by induction on $l$. Note that $\rho(Z')=\overline{\rho}(Z')=\frac{1}{2}$ by Lemma \ref{lemma4.2}.\\

Base case ($l=0$): This follows immediately from the definition: $$\mathcal{P}(Z',\Psi_1)=\rho(Z')=\frac{1}{2}. $$

 Induction step: Assume the claim holds for length $l=n-1$, i.e., $$\mathcal{P}((x_2,y_2)(x_3,y_3)\cdots (x_n,y_n)Z',\Psi_1)=\rho({\rm trim}(x_2x_3\cdots x_n)Z'). $$

Now consider length $l=n$, i.e., the configuration $(x_1,y_1)\alpha' Z'$ where $\alpha'=(x_2,y_2)\cdots (x_n,y_n)$. 

From the rules in (\ref{eq2}), we have $$(x_1,y_1)\alpha'Z\rightarrow^{\frac{1}{2}} X_{(x_1,y_1)}\alpha'Z'\rightarrow^{1}\alpha' Z'$$ and $$(x_1,y_1)\alpha'Z'\rightarrow^{\frac{1}{2}}\alpha' Z'.$$ There are $3$ cases:
\begin{enumerate}
\item { if $x_1=\bullet$, then
$$\aligned
(\bullet,y_1)\alpha'Z&\rightarrow^{\frac{1}{2}}X_{(\bullet,y_1)}\alpha'Z'\rightarrow^1\alpha'Z'\\
(\bullet,y_1)\alpha'Z&\rightarrow^{\frac{1}{2}}\alpha'Z'.
\endaligned$$
Thus,
$$\aligned
\mathcal{P}((x_1,y_1)\alpha'Z',\Psi_1)=&\frac{1}{2}\times \mathcal{P}(\alpha'Z',\Psi_1)+\frac{1}{2}\times \mathcal{P}(\alpha'Z',\Psi_1)\\
=&\mathcal{P}(\alpha'Z',\Psi_1)\\
=&\rho({\rm trim}(x_1x_2\cdots x_n)Z');
\endaligned$$
}
\item {if $x_1=B$, then
$$\aligned
(B,y_1)\alpha'Z&\rightarrow^{\frac{1}{2}}X_{(B,y_1)}\alpha'Z'\\
(B,y_1)\alpha'Z&\rightarrow^{\frac{1}{2}}\alpha'Z'.
\endaligned$$
Hence,
$$\aligned
\mathcal{P}((x_1,y_1)\alpha' Z',\Psi_1)=&\frac{1}{2}\times 0+\frac{1}{2}\times\mathcal{P}(\alpha'Z',\Psi_1)=\frac{1}{2}\times\rho({\rm trim}(x_2\cdots x_n)Z')\\
=&\rho({\rm trim}(x_1x_2\cdots x_n)Z');
\endaligned$$
}
\item {if $x_1=A$, then
$$\aligned
(A,y_1)\alpha'Z&\rightarrow^{\frac{1}{2}}X_{(A,y_1)}\alpha'Z'\\
(A,y_1)\alpha'Z&\rightarrow^{\frac{1}{2}}\alpha'Z'.
\endaligned$$
Hence,
$$\aligned
\mathcal{P}((x_1,y_1)\alpha'Z',\Psi_1)=&\frac{1}{2}+\frac{1}{2}\times\mathcal{P}(\alpha'Z',\Psi_1)\\
=&\frac{1}{2}+\frac{1}{2}\rho({\rm trim}(x_2\cdots x_n)Z')\\
=&\rho({\rm trim}(x_1x_2\cdots x_n)Z').
\endaligned$$
}
\end{enumerate}

In all cases, the following claim holds: $$\mathcal{P}(\{\pi\in\mbox{$RUN(F\alpha Z')$}\,:\,\pi\models^L\Psi_1\})=\rho(u'_{j_1}u'_{j_2}\cdots u'_{j_k}Z'). $$

The same reasoning applies symmetrically to $\Psi_2$, i.e., we have $$\mathcal{P}(\{\pi\in\mbox{$RUN(S\alpha Z')$}\,:\,\pi\models^L\Psi_2\})=\overline{\rho}(v'_{j_1}v'_{j_2}\cdots v'_{j_k}Z').$$
\end{proof}

Combining Lemma \ref{lemma4.2} and Lemma \ref{lemma4.3}, we obtain the following:
\begin{lemma}
\label{lemma4.4}
Let $\alpha = (u_{j_1},v_{j_1})(u_{j_2},v_{j_2})\cdots(u_{j_k},v_{j_k})\in\Sigma^*\times\Sigma^*$ be the sequence of pairs pushed onto the stack by $\Theta'$. Let $(u'_i,v'_i)$, $1\leq i\leq k$, be the pairs obtained after erasing all occurrences of $\bullet$ from $u_i$ and $v_i$. Then $u'_{j_1}\cdots u'_{j_k} = v'_{j_1}\cdots v'_{j_k}$
if and only if 
$$
\mathcal{P}(\{\pi\in\text{$RUN(F\alpha Z')$}\,:\,\pi\models^L\Psi_1\})+\mathcal{P}(\{\pi\in\text{$RUN(S\alpha Z')$}\,:\,\pi\models^L\Psi_2\}) =1.
$$\Q.E.D
\end{lemma}

With Lemma \ref{lemma4.4} in hand, we can prove the following:

\begin{lemma}
\label{lemma4.5}
Let $\alpha = (u_{j_1},v_{j_1})(u_{j_2},v_{j_2})\cdots(u_{j_k},v_{j_k})\in\Sigma^*\times\Sigma^*$ be the sequence of pairs pushed onto the stack by $\Theta'$, and let $(u'_i,v'_i)$, $1\leq i\leq j_k$, be the corresponding pairs after removing  all $\bullet$ symbols. Then, for any rational constant $t\in(0,1)\cap\mathbb{Q}$,
\begin{equation}
\label{eq4}
u'_{j_1}\cdots u'_{j_k} = v'_{j_1}\cdots v'_{j_k}
\end{equation}
if and only if $$\Theta', N\alpha Z'\models^L\mathcal{P}_{=\frac{t}{2}}(\Psi_1)\wedge\mathcal{P}_{=\frac{1-t}{2}}(\Psi_2).$$
\end{lemma}
\begin{proof}
When $\alpha$ is pushed onto the stack of $\Theta'$, the resulting configuration is $C\alpha Z'$ (read the stack from left to right). The only applicable rule is $C\rightarrow N$, so $C\alpha Z'$ transitions to $N\alpha Z'$ with probability $1$. 

The ``if" part. Suppose $\Theta',N\alpha Z'\models^L\mathcal{P}_{=\frac{t}{2}}(\Psi_1)\wedge\mathcal{P}_{=\frac{1-t}{2}}(\Psi_2)$.

Then the probability of paths from $N\alpha Z'$ satisfying $\Psi_1$ is $\frac{t}{2}$ and the probability of paths from $N\alpha Z'$ satisfying $\Psi_2$ is $\frac{1-t}{2}$. Since $\mathcal{P}(N\rightarrow F)=\mathcal{P}(N\rightarrow S)=\frac{1}{2}$, it follows that
$$
\mathcal{P}(\{\pi\in RUN(F\alpha Z'):\pi\models^L\Psi_1\})=t,\qquad\mathcal{P}(\{\pi\in RUN(S\alpha Z'):\pi\models^L\Psi_2\})=1-t.
$$
Adding these probabilities yields $1$, i.e., the following holds
\begin{equation}
\label{eq5}
\mathcal{P}(\{\pi\in\text{$RUN(F\alpha Z')$}\,:\,\pi\models^L\Psi_1\})+\mathcal{P}(\{\pi\in\text{$RUN(S\alpha Z')$}\,:\,\pi\models^L\Psi_2\})=t+(1-t) =1.
\end{equation}
By (\ref{eq5}) and Lemma \ref{lemma4.4}, equation (\ref{eq4}) holds.

The ``only if" part. Assume (\ref{eq4}) holds. Then, by Lemma \ref{lemma4.4},
$$
\mathcal{P}(\{\pi\in\text{$RUN(F\alpha Z')$}\,:\,\pi\models^L\Psi_1\})+\mathcal{P}(\{\pi\in\text{$RUN(S\alpha Z')$}\,:\,\pi\models^L\Psi_2\}) =1.
$$
Namely, $$\mathcal{P}(\{\pi\in\text{$RUN(F\alpha Z')$}\,:\,\pi\models^L\Psi_1\})=1-\mathcal{P}(\{\pi\in\text{$RUN(S\alpha Z')$}\,:\,\pi\models^L\Psi_2\}).$$ Combined with $\mathcal{P}(N\rightarrow F)=\mathcal{P}(N\rightarrow S)=\frac{1}{2}$, this implies $$\Theta',N\alpha Z' \models^L\mathcal{P}_{=\frac{t}{2}}(\Psi_1)\wedge\mathcal{P}_{=\frac{1-t}{2}}(\Psi_2).$$ The lemma follows.
\end{proof}

We are now ready to prove the following lemma:
\begin{lemma}
\label{lemma4.6}
Let $\pi$ be a path of the $\omega$-pBPA $\Theta'$, starting from $Z$ and reaching configuration $C\alpha Z'$, where $\alpha$ is guessed by $\Theta'$ as a candidate solution of the modified PCP instance. Then
\begin{equation}
 \label{eq6}
 \begin{split}
 \Theta',Z\models^L\mathcal{P}_{>0}({\bf true}{\bf U}[C\wedge \mathcal{P}_{=1}({\bf X}[\mathcal{P}_{=\frac{t}{2}}(\Psi_1)\wedge\mathcal{P}_{=\frac{1-t}{2}}(\Psi_2)])])
 \end{split}
 \end{equation}
 if and only if $\alpha$ is a solution of the modified PCP instance (for any constant $t\in(0,1)\cap\mathbb{Q}$).
\end{lemma}
\begin{proof}
The following equivalences hold:\\
(\ref{eq4} ) is true
$$\aligned
&\Leftrightarrow\,\,\Theta',N\alpha Z'\models^L\mathcal{P}_{=\frac{t}{2}}(\Psi_1)\wedge\mathcal{P}_{=\frac{1-t}{2}}(\Psi_2)\qquad\text{ (by Lemma \ref{lemma4.5})}\\
&\Leftrightarrow\,\,\Theta',C\alpha Z'\models^L{\bf X}[\mathcal{P}_{=\frac{t}{2}}(\Psi_1)\wedge\mathcal{P}_{=\frac{1-t}{2}}(\Psi_2)]\qquad\text{(by $C\rightarrow N$)}\\
&\Leftrightarrow\,\,\Theta',C\models^L\mathcal{P}_{=1}({\bf X}[\mathcal{P}_{=\frac{t}{2}}(\Psi_1)\wedge\mathcal{P}_{=\frac{1-t}{2}}(\Psi_2)])\qquad\text{(by $\mathcal{P}(C\rightarrow N)=1$)}\\
 &\Leftrightarrow\,\,\Theta',Z\models^L\mathcal{P}_{>0}({\bf true}{\bf U}[C\wedge \mathcal{P}_{=1}({\bf X}[\mathcal{P}_{=\frac{t}{2}}(\Psi_1)\wedge\mathcal{P}_{=\frac{1-t}{2}}(\Psi_2)])])\qquad\text{(by Lemma \ref{lemma4.1})}
\endaligned$$

Thus $$\Theta',Z\models^L\mathcal{P}_{>0}({\bf true}{\bf U} [C\wedge \mathcal{P}_{=1}({\bf X}[\mathcal{P}_{=\frac{t}{2}}(\Psi_1)\wedge\mathcal{P}_{=\frac{1-t}{2}}(\Psi_2)])])$$ if and only if $\alpha$ is a solution of the modified PCP instance.
\end{proof}

The formula $$\mathcal{P}_{>0}({\bf true}{\bf U}[C\wedge \mathcal{P}_{=1}({\bf X}[\mathcal{P}_{=\frac{t}{2}}(\Psi_1)\wedge\mathcal{P}_{=\frac{1-t}{2}}(\Psi_2)])])$$ is a strict $\omega$-PCTL formula because $\Psi_1$ and $\Psi_2$ are $\omega$-PCTL path formulas with Büchi formula.

\subsection{Proof of Theorem \ref{theorem1}}
By Lemma \ref{lemma4.6}, formula (\ref{eq6}) holds if and only if $\alpha$ is a solution of the modified PCP problem. Therefore, any algorithm deciding whether formula (\ref{eq6}) holds could be used to decide the modified Post Correspondence Problem, which is undecidable. This completes the proof of Theorem \ref{theorem1}. \Q.E.D

\begin{remark}
\label{remark4.1}
Note that the constant $t$ in (\ref{eq6}) can be any rational number in $(0,1)\cap\mathbb{Q}$. For concreteness, one may fix $t=0.3$ in formula (\ref{eq6}), yielding the concrete formula  $$\mathcal{P}_{>0}({\bf true}{\bf U}[C\wedge \mathcal{P}_{=1}({\bf X}[\mathcal{P}_{=\frac{0.3}{2}}(\Psi_1)\wedge\mathcal{P}_{=\frac{0.7}{2}}(\Psi_2)])]). $$
\end{remark}

\begin{remark}
\label{remark4.2}
 Corollary \ref{corollary2} follows immediately since $\omega$-PCTL is a sublogic of $\omega$-PCTL$^*$. For Corollary \ref{corollary3}, it suffices to replace the rule $Z'\rightarrow Z'$ by $(q,Z')\rightarrow (q,Z')$ for some state $q\in F_{\rm inal}$ (turning the system into an $\omega$-pPDS) and replace the sub-formula $$Z'\wedge\mathcal{P}_{=1}\Big({\rm B\text{ü}chi}(Z')\Big)$$ in (\ref{eq3}) by $$(q,Z')\wedge\mathcal{P}_{=1}\Big({\rm B\text{ü}chi}((q,Z'))\Big). $$
\end{remark}

\begin{remark}
Careful readers may notice that PCTL is a sublogic of $\omega$-PCTL, and since the problem of model-Checking of $\omega$-pBPA against PCTL is undecidable \cite{LL24}, the result for $\omega$-PCTL follows immediately. This observation is correct. However, the contribution of this section lies in showing undecidability specifically for {\em strict} (hard) $\omega$-PCTL formulas, i.e., those that are not PCTL formulas.
\end{remark}

\begin{remark}
Some readers might wonder whether the PCTL$^*$ formula used in \cite{BBFK14} to prove the undecidability for {\em stateless} pPDS against PCTL$^*$ already falls into the $\omega$-PCTL fragment, and how it differs from ours. While we are not certain about the former, we note that our formula (\ref{eq6}) combines the PCTL formula (10) from \cite{LL24} (used to show the undecidability of PCTL model checking for pBPA, which was left open in \cite{BBFK14}) with the Büchi formula $\mathcal{P}_{=1}({\rm B\text{ü}chi}(Z'))$. For a detailed comparison between formula (10) of \cite{LL24} and the formulas used in \cite{BBFK14}, we refer the reader to \cite{LL24}.
\end{remark}

\section{Lower Bound for Model-Checking $\omega$-pBPA against $\omega$-bPCTL}
\label{sec:proof_of_theorem2}

\subsection{$\omega$-Bounded Probabilistic Computational Tree Logic}

We first introduce $\omega$-bounded probabilistic computational tree logic ($\omega$-bPCTL). 

The logic $\omega$-bounded PCTL ($\omega$-bPCTL) is obtained by replacing until operator ${\bf U}$ in $\omega$-PCTL (defined in  Subsection \ref{subsec:omega-pctl}) with the bounded until operator ${\bf U}^{\leq k}$.

Let $AP$ be a fixed set of atomic propositions. 

Formally, the syntax of {\em $\omega$-bounded probabilistic computational tree logic} ($\omega$-bPCTL) is defined as follows:
$$\aligned
 \Phi&::={\bf true}\text{ $|$ }p\text{ $|$ }\neg\Phi\text{ $|$ }\Phi_1\wedge\Phi_2\text{ $|$ }\mathcal{P}_{\bowtie r}(\varphi)\\
 \varphi&::={\bf X}\Phi\text{ $|$ } \Phi_1{\bf U}^{\leq k}\Phi_2\text{ $|$ }\varphi^{\omega}\\
 \varphi^{\omega}&::={\rm B\text{ü}chi}(\Phi)\text{ $|$ }{\rm coB\text{ü}chi}(\Phi)\text{ $|$ }\varphi_1^{\omega}\wedge\varphi_2^{\omega}\text{ $|$ }\varphi_1^{\omega}\vee\varphi_2^{\omega},
\endaligned$$
where $\Phi$ is a state formula, $\varphi$ is a path formula, and $\varphi^{\omega}$ denotes an infinitary path formula (i.e., one that depends on the set of states visited infinitely often along a path). Here, $p \in AP$ is an atomic proposition, $\bowtie\in\{>, \leq,>,\geq\}$, and $r\in[0,1]\cap\mathbb{Q}$ is a rational constant.

The semantics of $\omega$-bPCTL over a Markov chain $\mathcal{M}$ is defined similarly to that of $\omega$-PCTL in Subsection \ref{subsec:omega-pctl}, except for the bounded until operator, which is interpreted as follows: 
$$\mathcal{M},\pi\models^L\Phi_1{\bf U}^{\leq k}\Phi_2\quad{\rm iff}\quad\text{$\exists 0\leq i\leq k$  such that $\mathcal{M},\pi[i]\models^L\Phi_2$ and $\forall j<i$, $\mathcal{M},\pi[j]\models^{L}\Phi_1$.} $$

\subsection{Bounded Post Correspondence Problem}

Formally, a bounded PCP instance consists of a finite alphabet $\Sigma$ and a finite set $\{(u_i,v_i)\,:\,1\leq i\leq n\}\subseteq\Sigma^*\times\Sigma^*$ of $n$ pairs of strings over $\Sigma$, together with a positive integer $K\leq n$. The problem asks whether there exists a word $j_1j_2\cdots j_k\in\{1,2,\cdots,n\}^+$ with $k\leq K$ such that $u_{j_1}u_{j_2}\cdots u_{j_k}=v_{j_1}v_{j_2}\cdots v_{j_k}$.

The bounded version of the {\it Post Correspondence Problem} is $\mathit{NP}$-complete; see e.g., \cite{GJ79}, page 228.
\begin{theoremsec}[\cite{GJ79}, p. 228]
\label{theorem5.1}
The {\it bounded Post Correspondence Problem} is $\mathit{NP}$-complete.\Q.E.D
\end{theoremsec}

Likewise, we will define a modified version of the bounded PCP, which is more convenient for our purpose. Since each word $w\in\Sigma^*$ is of finite length, let $m=\max\{|u_i|,|v_i|\}_{1\leq i\leq n}$.

We pad each string $u_i$ and $v_i$ with the special symbol ``$\bullet$" (possibly at the beginning or end) to obtain strings $u'_i$ and $v'_i$ of length exactly $m$. The modified bounded PCP then asks whether there exists a sequence $j_1j_2\cdots j_k\in\{1,2,\cdots,n\}^+$ with $k\leq K$ such that $u'_{j_1}\cdots u'_{j_k}=v'_{j_1}\cdots v'_{j_k}$ after removing all occurrences of $\bullet$.

It is easy to see that this modified version is equivalent to the original bounded PCP. Hence, it is also $\mathit{NP}$-complete:
\begin{theoremsec}
\label{theorem5.2}
The {\it modified bounded Post Correspondence Problem} is $\mathit{NP}$-complete.\Q.E.D
\end{theoremsec}

\subsection{Proofs of Technical Lemmas}

Obviously, an arbitrary instance of the modified Post Correspondence Problem (see Subsection \ref{sec:post_correspondence_problem}) cannot be encoded into an $\omega$-bPCTL formula, because only modified PCP instances with length at most $\leq k$ (i.e., the modified bounded PCP instance) can be expressed using the bounded until operator $\Phi_1{\bf U}^{\leq k}\Phi_2$ of $\omega$-bPCTL. Thus, we have the following (which is immediate from the above discussion):

\begin{theorem}
\label{theorem5}
Model checking stateless probabilistic $\omega$-pushdown systems ($\omega$-pBPA) against $\omega$-bounded probabilistic computational tree logic ($\omega$-bPCTL) is decidable.
\end{theorem}

The proof of the above theorem is straightforwad and therefore omitted. As already discussed above, only modified PCP instances whose length at most $k$ (i.e., modified bounded PCP instances) can be expressed using the path formula $\Phi_1{\bf U}^{\leq k}\Phi_2$ of $\omega$-PCTL. Consequently, arbitrary (unbounded) modified PCP instances cannot be encoded into $\omega$-bPCTL formulas, which implies that the model-checking problem is decidable. The proof of Theorem \ref{theorem2} further confirms this.

To prove Theorem \ref{theorem2}, we reduce the bounded Post Correspondence Problem (or equivalently, the modified bounded Post Correspondence Problem) to the model-checking problem for stateless probabilistic $\omega$-pushdown systems against $\omega$-bPCTL. We prove Theorem \ref{theorem2} by a reduction from the {\it modified bounded Post Correspondence Problem}. However, the construction is considerably more involved than the one presented in Section \ref{sec:proof_of_theorem_1}. 

We fix $\Sigma = \{A, B, \bullet\}$, and the stack alphabet $\Gamma$ of the $\omega$-$pBPA$ be defined as follows:\footnote{Note that $m$ is the common length of the padded string $u'_i$ and $v'_i$ (i.e., $m=\max\{|u_i|,|v_i|\}_{1\leq i\leq n}$), and $n$ is the number of pairs (i.e., $n=|\{u_i,v_i\}_{1\leq i\leq n}|$).}
$$\aligned
\Gamma=&\big\{Z\big\}\cup\big\{1,2,\cdots,n\big\}\cup\big\{Z'\big\}\cup\big\{G_{l_k,k}^j\,:\,1\leq k\leq n,1\leq l_k\leq n,1\leq j\leq m+1\big\}\\
&\cup\big\{(x,y),X_{(x,y)}\,:\,(x,y)\in\Sigma\big\}\cup\big\{C,F,S,N\big\}\\
\endaligned$$

The elements of $\Gamma$ also serve as atomic propositions. We now describe how to construct the desired {\em stateless probabilistic $\omega$-pushdown system} $$\triangle=(\Gamma, \delta,Z,F_{\rm inal}=\{Z'\},{\bf P}, \mathcal{F}). $$

Our $\omega$-$pBPA$ $\triangle$ operates in two phases (this is similar to the construction in Section \ref{sec:proof_of_theorem_1}). The first phase guesses a potential solution to a modified bounded PCP instance by pushing pairs of words $(u_i,v_i)$ onto the stack, according to the following transition rules:

\begin{equation}
\label{eq7}
\begin{split}
Z\rightarrow&1Z'\,|\,2Z'\,|\cdots|\,nZ';\quad\text{(with $\mathcal{P}(Z\rightarrow kZ')=\frac{1}{n}$ for all $k\in[n]$)}\\
k\rightarrow&G^1_{1,k}\,|\,G^1_{2,k}\,|\cdots\,|\,G_{n,k}^1;\quad\text{(with $\mathcal{P}(k\rightarrow G^1_{l_k,k})=\frac{1}{n}$ for $l_k\in[n]$)}\\ 
G_{l_k,k}^j\rightarrow& G_{l_k,k}^{j+1}(u_{l_k}(j),v_{l_k}(j));\quad\text{(with $\mathcal{P}(G^j_{l_k,k}\rightarrow G_{l_k,k}^{j+1}(u_{l_k}(j),v_{l_k}(j)))=1$ for $j\in[m]$)}\\
G_{l_k,k}^{m+1}\rightarrow& G_{1,k-1}^1\,|\,G_{2,k-1}^1\,|\cdots|G_{n,k-1}^1;\quad\text{(with $\mathcal{P}(G_{l_k,k}^{m+1}\rightarrow G_{l_{k-1},k-1}^1)=\frac{1}{n}$ for $l_{k-1}\in[n]$)}\\
G_{l_{k-1},k-1}^j\rightarrow&G_{l_{k-1},k-1}^{j+1}(u_{l_{k-1}}(j),v_{l_{k-1}}(j));\quad\text{(with $\mathcal{P}(G^j_{l_{k-1},k-1}\rightarrow G_{l_{k-1},k-1}^{j+1}(u_{l_{k-1}}(j),v_{l_{k-1}}(j)))=1$}\\
&\qquad\qquad\qquad\qquad\qquad\text{for $j\in[m]$)}\\
G_{l_{k-1},k-1}^{m+1}\rightarrow& G_{1,k-2}^1\,|\,G_{2,k-2}^1\,|\cdots\,|\,G_{n,k-2}^1;\quad\quad\text{(with $\mathcal{P}(G_{l_{k-1},k-1}^{m+1}\rightarrow G_{l_{k-2},k-2}^1)=\frac{1}{n}$ for $l_{k-2}\in[n]$)}\\
\vdots&\\
\vdots&\\
G_{l_2,2}^{m+1}\rightarrow&G_{1,1}^1\,|\,G_{2,1}^1\,|\cdots|\,G_{n,1}^1;\quad\quad\text{(with $\mathcal{P}(G_{l_2,2}^{m+1}\rightarrow G_{l_1,1}^1)=\frac{1}{n}$ for $l_2\in[n]$ and $l_1\in[n]$)}\\
G_{l_1,1}^j\rightarrow&G_{l_1,1}^{j+1}(u_{l_1}(j),v_{l_1}(j));\quad\text{(with $\mathcal{P}(G_{l_1,1}^j\rightarrow G_{l_1,1}^{j+1}(u_{l_1}(j),v_{l_1}(j))=1$ for $j\in[m]$) }\\
G_{l_1,1}^{m+1}\rightarrow& C\quad\text{(with $\mathcal{P}(G_{l_1,1}^{m+1}\rightarrow C)=1$)}.
\end{split}
\end{equation}

In the above probabilistic transition rules (\ref{eq7}), we assume that $k$ is the guessed bound chosen by $\triangle$. Clearly, $k$ can be any positive integer in $[n]$.

In the above rules, we first note that $(u_i,v_i)=(u_i(1)u_i(2)\cdots u_i(m),v_i(1)v_i(2)\cdots v_i(m))$. Thus, $(u_i(j),v_i(j))$ means that, after selecting the $i$-th pair $(u_i,v_i)$, we further selecting the $j$-th symbol in $u_i$ (say $x'$) and the $j$-th symbol in $v_i$ (say $y'$) to form the pair $(x',y')$. Obviously, the symbol $Z$ serves as the initial stack symbol. 

It process begins by guessing a bound $k$ (where $k\leq n$) for an instance of the modified bounded Post Correspondence Problem and pushing the string $kZ'$ ($\in\Gamma^*$) onto the stack with probability $\frac{1}{n}$. The symbol at the top of the stack is then $k$ (the stack is read from left to right). According to the rules in (\ref{eq7}), this guessed bound $k$ is replaced, with probability $\frac{1}{n}$, by $G^1_{l_k,k}$, where $l_k\in[n]$.

Next, the symbol at the top of the stack is $G_{l_k,k}^1$. The rules in (\ref{eq7}) state that $G_{l_k,k}^1$ is replaced with probability $1$ by $G_{l_k,k}^2(u_{l_k}(1),v_{l_k}(1))$. This process repeats until $G_{l_k,k}^{m+1}(u_{l_k}(m),v_{l_k}(m))$ reaches the top of the stack, indicating that the $k$-th pair of $(u_{l_k},v_{l_k})$ has been fully pushed onto the stack of $\triangle$. 

Then, with probability $\frac{1}{n}$, $\triangle$ replace the symbol $G_{l_k,k}^{m+1}$ by $G_{l_{k-1},k-1}^1$ (where $l_{k-1}\in[n]$), which initiates the pushing of the $(k-1)$-th pair $(u_{l_{k-1}},v_{l_{k-1}})$ onto the top of the stack.

The above process is repeated until the first pair $(u_{l_1},v_{l_1})$ (where $l_1\in[n]$) has been pushed onto the stack. This procedure produces a word $l_kl_{k-1}\cdots l_1\in\{1,2,\cdots,n\}^+$ with $k\leq n$ (where $l_1\in[n], l_2\in[n],\cdots,l_k\in[n]$), corresponding to the sequence of pairs $(u_{l_k},v_{l_k})(u_{l_{k-1}},v_{l_{k-1}})\cdots (u_{l_1},v_{l_1})$ that have been pushed onto the stack in that order. This sequence of the words $(u_{l_k},v_{l_k})(u_{l_{k-1}},v_{l_{k-1}})\cdots (u_{l_1},v_{l_1})$ represents a guessed candidate solution to a modified bounded PCP instance.

After this, with probability $1$, $\triangle$  pushes the symbol $C$ onto the stack, indicating that it will next check whether the sequence of pairs stored in the stack constitutes a solution to the modified bounded PCP instance. 

There are no other transition rules for $\triangle$ in the guessing phase besides those given in (\ref{eq7}). From the arguments above, we have the following lemma:

\begin{lemma}
\label{lemma5.1}
A configuration of the form $C\alpha Z'$ with $|\alpha|\leq nm$ is reachable from $Z$ if and only if $\alpha\equiv(x_1,y_1)\cdots(x_t,y_t)$ where $x_i,y_i\in\Sigma$ ($1\leq i\leq t$), and there exists a word $l_kl_{k-1}\cdots l_1\in\{1,2,\cdots,n\}^+$ with $k\leq n$ such that $x_1\cdots x_t=u_{l_1}\cdots u_{l_k}$ and $y_1\cdots y_t = v_{l_1}\cdots v_{l_k}$ (reading the stack from left to right).\footnote{Note that, by (\ref{eq7}), the pair $(u_{l_k},v_{l_k})$ is the first one pushed onto the stack, followed by $(u_{l_{k-1}},v_{l_{k-1}})$, $\cdots$, and finally $(u_{l_1},v_{l_1})$. Also note that the stack is read from left to right, i.e., the top of the stack is on the left.} The probability $p$ of reaching $C\alpha Z'$ from $Z$ satisfies $p>0$.\Q.E.D
\end{lemma}

The next step is for $\triangle$ to verify the stored sequence of word pairs. The corresponding transition rules are given as follows:
\begin{equation}
\label{eq8}
\begin{split}
C\rightarrow& N,\quad\text{(with $\mathcal{P}(C\rightarrow N)=1$)}\\
N\rightarrow& F\,|\,S,\quad\text{(with $\mathcal{P}(N\rightarrow F)=\mathcal{P}(N\rightarrow S)=\frac{1}{2}$)}\\
 F\rightarrow&\epsilon,\quad\text{(with $\mathcal{P}(F\rightarrow \epsilon)=1$)}\\
 S\rightarrow&\epsilon,\quad\text{(with $\mathcal{P}(S\rightarrow \epsilon)=1$)}\\
 (x,y)\rightarrow& X_{(x,y)}\,|\,\epsilon,\quad\text{(with $\mathcal{P}((x,y)\rightarrow X_{(x,y)})=\frac{1}{2}$ and $\mathcal{P}((x,y)\rightarrow \epsilon)=\frac{1}{2}$)}\\
  X_{(x,y)}\rightarrow&\epsilon,\quad\text{(with $\mathcal{P}(X_{(x,y)}\rightarrow \epsilon)=1$)}\\
    Z'\rightarrow& Z',\quad\text{(with $\mathcal{P}(Z'\rightarrow Z')=1$)} 
  \end{split}
\end{equation}

The Büchi accepting condition of $\triangle$ is that, for any run $r=(Z)(\gamma_2)(\gamma_3)\cdots$, $r$ is successful iff ${\rm Inf}(r)\cap F_{\rm inal}\neq\emptyset$.

\begin{remark}
\label{remark5.1}
We emphasize that there are no other rules in the verifying step besides those described in (\ref{eq8}).
\end{remark}

When the stack symbol $C$ is at the top of the stack, $\triangle$ checks whether the previously guessed sequence is a solution to the modified bounded PCP instance. It first replaces $C$ with $N$ at the top of the stack with probability $1$. Then it replaces $N$ by $F$ or $S$, each with probability $$\mathcal{P}(N\rightarrow F)=\mathcal{P}(N\rightarrow S)=\frac{1}{2},$$ depending on whether $\triangle$ chooses to check the $u$'s or $v$'s. 

We stress that Lemma \ref{lemma4.2} remains applicable in this section. For readability, we restate it below:
\begin{lemma}[Lemma \ref{lemma4.2} in Section \ref{sec:proof_of_theorem_1}]
\label{lemma5.2}
Let $\vartheta$ and $\overline{\vartheta}$ be two functions from $\{A,B,Z'\}$ to $\{0,1\}$, defined by
$$
             \vartheta(x)=\left\{
                              \begin{array}{ll}
                                1, & \hbox{$X=Z'$;} \\
                                1, & \hbox{$X=A$;} \\
                                0, & \hbox{$X=B$.}
                              \end{array}
                            \right.\,\,\,
             \overline{\vartheta}(x)=\left\{
                                         \begin{array}{ll}
                                           1, & \hbox{$X=Z'$;} \\
                                           0, & \hbox{$X=A$;} \\
                                           1, & \hbox{$X=B$.}
                                         \end{array}
                                       \right.
$$

Further, let $\rho$ and $\overline{\rho}$ be two functions from $\{A,B\}^+Z'$ to $[0,1]$, given by
$$
 \rho(x_1x_2\cdots x_n)\overset{\rm def}{=}\sum_{i=1}^n\vartheta(x_i)\frac{1}{2^i},\quad
 \overline{\rho}(x_1x_2\cdots x_n)\overset{\rm def}{=}\sum_{i=1}^n\overline{\vartheta}(x_i)\frac{1}{2^i}.
$$

Then, for any $(u'_{j_1},v'_{j_1}),(u'_{j_2},v'_{j_2}),\cdots,(u'_{j_k},v'_{j_k})\in\{A,B\}^+\times\{A,B\}^+$,
$$
u'_{j_1}u'_{j_2}\cdots u'_{j_k} = v'_{j_1}v'_{j_2}\cdots v'_{j_k}
$$
if and only if
$$
\rho(u'_{j_1}\cdots u'_{j_k}Z')+\overline{\rho}(v'_{j_1}v'_{j_2}\cdots v'_{j_k}Z') = 1
$$
\end{lemma}
\begin{proof}
See the proof of Lemma \ref{lemma4.2}.
\end{proof}

 Because of Lemma \ref{lemma5.2}, we define two $\omega$-bPCTL path formulas $\varphi_3$ and $\varphi_4$, which will also be useful, as follows:
\begin{equation}
\label{eq9}
\begin{split}
\varphi_3 =& (\neg S\wedge\bigwedge_{z\in\Sigma}\neg X_{(B,z)}){\bf U}^{\leq 2nm}\left(\Big[\bigvee_{z\in\Sigma}X_{(A,z)}\Big]\vee \Big[Z'\wedge\mathcal{P}_{=1}\Big({\rm B\text{ü}chi}(Z')\Big)\Big]\right),\\
\varphi_4 =& (\neg F\wedge\bigwedge_{z\in\Sigma}\neg X_{(z,A)}){\bf U}^{\leq 2nm}\left(\Big[\bigvee_{z\in\Sigma}X_{(z,B)}\Big]\vee \Big[Z'\wedge\mathcal{P}_{=1}\Big({\rm B\text{ü}chi}(Z')\Big)\Big]\right).
\end{split}
\end{equation}

These two path formulas are related to $\rho(u'_{j_1}\cdots u'_{j_k}Z')$ and $\overline{\rho}(v'_{j_1}\cdots v'_{j_k}Z')$, respectively. We prove this connection in Lemma \ref{lemma5.3} below.

Recall that ${\rm trim}(b_1b_2\cdots b_n)$ denotes the word $\in\{A,B\}^*$ obtained by erasing all occurrences of ``$\bullet$" in $b_1b_2\cdots b_n$. Similarly, ${\rm trim}(b_2b_3\cdots b_n)$ is the word $\in\{A,B\}^*$ obtained by erasing all ``$\bullet$" symbols in $b_2b_3\cdots b_n$. We then have the following lemma.

\begin{lemma}
\label{lemma5.3}
Let $\alpha$ be the sequence of word pairs pushed onto the stack by $\triangle$, where $\alpha = (u_{l_1},v_{l_1})(u_{l_2},v_{l_2})\cdots (u_{l_k},v_{l_k})\in\Sigma^*\times\Sigma^*$ with $|\alpha|\leq nm$\footnote{Note that $|u_{l_1}u_{l_2}\cdots u_{l_k}|=|v_{l_1}v_{l_2}\cdots v_{l_k}|$, thus $|\alpha|$ is defined as $|u_{l_1}u_{l_2}\cdots u_{l_k}|$.}, and let $(u'_{l_i},v'_{l_i})$, $1\leq i\leq k$, be the pairs obtained after erasing all $\bullet$ symbols from $(u_{l_i},v_{l_i})$. Then
$$\aligned
\mathcal{P}(\{\pi\in\text{$RUN(F\alpha Z')$}\,:\,\pi\models^{L}\varphi_3\})=&\rho(u'_{l_1}u'_{l_2}\cdots u'_{l_k}Z')\\
\mathcal{P}(\{\pi\in\text{$RUN(S\alpha Z')$}\,:\,\pi\models^{L}\varphi_4\})=&\overline{\rho}(v'_{l_1}v'_{l_2}\cdots v'_{l_k}Z'),
\endaligned$$where $\varphi_3$ and $\varphi_4$ are defined in (\ref{eq9}).
\end{lemma}
\begin{proof}
Let 
$$\aligned
\mathcal{P}(F\alpha Z',\varphi_3)\overset{\rm def}{=}&\mathcal{P}(\{\pi\in\text{$RUN(F\alpha Z')$}\,:\,\pi\models^{L}\varphi_3\}),\\
\mathcal{P}(S\alpha Z',\varphi_4)\overset{\rm def}{=}&\mathcal{P}(\{\pi\in\text{$RUN(S\alpha Z')$}\,:\,\pi\models^{L}\varphi_4\}).
\endaligned$$ 
Suppose $$x_1x_2\cdots x_l=u_{l_1}u_{l_2}\cdots u_{l_k}$$ and $$y_1y_2\cdots y_l=v_{l_1}v_{l_2}\cdots v_{l_k}; $$ We prove by induction on $l$ (i.e., the length of $\alpha$) that $$\mathcal{P}(F\alpha Z',\varphi_3)=\rho({\rm trim}(x_1x_2\cdots x_l)Z');$$ the proof for

$$\mathcal{P}(S\alpha Z',\varphi_4)=\overline{\rho}({\rm trim}(y_1y_2\cdots y_l)Z')$$ is analogous.

Note that by (\ref{eq8}), $F\alpha Z'\rightarrow\alpha Z'$ with probability $1$. Hence $$\mathcal{P}(F\alpha Z',\varphi_3)=\mathcal{P}(\alpha Z',\varphi_3).$$ It therefore suffices to show that $\mathcal{P}(\alpha Z',\varphi_3)=\rho({\rm trim}(x_1x_2\cdots x_l)Z')$.
We proceed by induction on $l$. First, recall from Lemma \ref{lemma5.2} that $\rho(Z')=\overline{\rho}(Z')=\frac{1}{2}$.\\

Base case ($l=0$): This follows immediately from the definition, i.e., $$\mathcal{P}(Z',\varphi_3)=\rho(Z')=\frac{1}{2}. $$

 Induction step: Assume the claim holds for $l=z-1$, i.e., $$\mathcal{P}((x_2,y_2)(x_3,y_3)\cdots (x_z,y_z)Z',\varphi_3)=\rho({\rm trim}(x_2x_3\cdots x_z)Z'). $$

Now consider the case of $l=z$, i.e., $\mathcal{P}((x_1,y_1)\alpha' Z',\varphi_3)$ where $\alpha'=(x_2,y_2)\cdots (x_z,y_z)$. 

By (\ref{eq8}), we have the transitions
$$(x_1,y_1)\alpha'Z\rightarrow^{\frac{1}{2}}X_{(x_1,y_1)}\alpha'Z'\rightarrow^{1}\alpha' Z'\quad\text{ and}\quad(x_1,y_1)\alpha'Z'\rightarrow^{\frac{1}{2}}\alpha' Z'.$$ We consider the following $3$ cases:
\begin{enumerate}
\item { if $x_1=\bullet$, then
$$\aligned
(\bullet,y_1)\alpha'Z&\rightarrow^{\frac{1}{2}}X_{(\bullet,y_1)}\alpha'Z'\rightarrow^{1}\alpha'Z'\\
(\bullet,y_1)\alpha'Z&\rightarrow^{\frac{1}{2}}\alpha'Z'.
\endaligned$$
Thus,
$$\aligned
\mathcal{P}((x_1,y_1)\alpha'Z',\varphi_3)=&\frac{1}{2}\times\mathcal{P}(\alpha'Z',\varphi_3)+\frac{1}{2}\times\mathcal{P}(\alpha'Z',\varphi_3)\\
=&\rho({\rm trim}(x_1x_2\cdots x_z)Z');
\endaligned$$
}
\item {if $x_1=B$, then
$$\aligned
(B,y_1)\alpha'Z&\rightarrow^{\frac{1}{2}}X_{(B,y_1)}\alpha'Z'\\
(B,y_1)\alpha'Z&\rightarrow^{\frac{1}{2}}\alpha'Z'.
\endaligned$$
Hence, 
$$\aligned
\mathcal{P}((x_1,y_1)\alpha' Z',\varphi_3)=&\frac{1}{2}\times 0+\frac{1}{2}\times\mathcal{P}(\alpha'Z',\varphi_3)\\
=&\frac{1}{2}\times\rho({\rm trim}(x_2\cdots x_n)Z')\\
=&\rho({\rm trim}(x_1x_2\cdots x_z)Z');
\endaligned$$
}
\item {if $x_1=A$, then
$$\aligned
(A,y_1)\alpha'Z&\rightarrow^{\frac{1}{2}}X_{(A,y_1)}\alpha'Z'\\
(A,y_1)\alpha'Z&\rightarrow^{\frac{1}{2}}\alpha'Z'.
\endaligned$$
Hence,
$$\aligned
\mathcal{P}((x_1,y_1)\alpha'Z',\varphi_3)=&\frac{1}{2}+\frac{1}{2}\times\mathcal{P}(\alpha'Z',\varphi_3)\\
=&\frac{1}{2}+\frac{1}{2}\rho({\rm trim}(x_2\cdots x_z)Z')\\
=&\rho({\rm trim}(x_1x_2\cdots x_z)Z').
\endaligned$$
}
\end{enumerate}

In all three cases, we obtain $\mathcal{P}((x_1,y_1)\alpha'Z',\varphi_3)=\rho({\rm trim}(x_1x_2\cdots x_z)Z')$. Therefore,
 $$\mathcal{P}(\{\pi\in\mbox{$RUN(F\alpha Z')$}\,:\,\pi\models^{L}\varphi_3\})=\rho(u'_{l_1}u'_{l_2}\cdots u'_{l_k}Z'). $$

The proof for $\mathcal{P}(\{\pi\in\mbox{$RUN(S\alpha Z')$}\,:\,\pi\models^{L}\varphi_4\})=\overline{\rho}(v'_{l_1}v'_{l_2}\cdots v'_{l_k}Z')$ is analogous.
\end{proof}

Combining Lemma \ref{lemma5.2} and Lemma \ref{lemma5.3}, we obtain the following result:
\begin{lemma}
\label{lemma5.4}
Let $\alpha = (u_{l_1},v_{l_1})(u_{l_2},v_{l_2})\cdots(u_{l_k},v_{l_k})\in\Sigma^*\times\Sigma^* $ with $|\alpha|\leq nm $ (reading the stack from left to right) be sequence of word pair pushed onto the stack by $\triangle$. Let $(u'_{l_i},v'_{l_i})$, for $1\leq i\leq k$, be the pairs obtained after erasing all $\bullet$ symbols from $u_{l_i}$ and $v_{l_i}$. Then $$u'_{l_1}\cdots u'_{l_k} = v'_{l_1}\cdots v'_{l_k} $$ if and only if $$\mathcal{P}(\{\pi\in\text{$RUN(F\alpha Z')$}\,:\,\pi\models^{L}\varphi_3\})+\mathcal{P}(\{\pi\in\text{$RUN(S\alpha Z')$}\,:\,\pi\models^{L}\varphi_4\}) =1. $$\Q.E.D
\end{lemma}

With the above lemma, we can further show the following:

\begin{lemma}
\label{lemma5.5}
Let $\alpha = (u_{l_1},v_{l_1})(u_{l_2},v_{l_2})\cdots(u_{l_k},v_{l_k})\in\Sigma^*\times\Sigma^*$ with $|\alpha|\leq nm$ (reading the stack from left to right) be the sequence of word pairs pushed onto the stack by $\triangle$. Let $(u'_{l_i},v'_{l_i})$, $1\leq i\leq k$, be the pairs obtained after erasing all $\bullet$ symbols from $u_{l_i}$ and $v_{l_i}$. Then
\begin{equation}
\label{eq10}
\begin{split}
u'_{l_1}\cdots u'_{l_k} = v'_{l_1}\cdots v'_{l_k}
\end{split}
\end{equation}
if and only if $$\widehat{M_{\triangle}}, N\alpha Z'\models^{L}\mathcal{P}_{=\frac{t}{2}}(\varphi_3)\wedge\mathcal{P}_{\frac{1-t}{2}}(\varphi_4),$$ where $t$ is any rational constant in $(0,1)\cap\mathbb{Q}$.
\end{lemma}

\begin{proof}
It is obvious that when $\alpha$ is pushed onto the stack of $\triangle$, the stack content is $C\alpha Z'$ (read from left to right). By (\ref{eq8}), the only applicable rule is $C\rightarrow N$. Thus, the stack content becomes $N\alpha Z'$ with probability $1$.

The ``if" part. Suppose that $\widehat{M_{\triangle}},N\alpha Z'\models^{L}\mathcal{P}_{=\frac{t}{2}}(\varphi_3)\wedge\mathcal{P}_{=\frac{1-t}{2}}(\varphi_4)$.

Then the probability of paths starting from $N$ that satisfy $\varphi_3$ is $\frac{t}{2}$, and the probability of paths from $N$ that satisfy $\varphi_4$ is $\frac{1-t}{2}$. Consequently, the probability of paths starting from $F$ that satisfy $\varphi_3$ is $t$, and the probability of paths starting from $S$ that satisfy $\varphi_4$ is $1 - t$. Since $\mathcal{P}(N\rightarrow F) =\frac{1}{2}$ and $\mathcal{P}(N \rightarrow S)= \frac{1}{2}$ (see (\ref{eq8})), it follows that
\begin{equation}
\label{eq11}
\mathcal{P}(\{\pi\in\text{$RUN(F\alpha Z')$}\,:\,\pi\models^{L}\varphi_3\})+\mathcal{P}(\{\pi\in\text{$RUN(S\alpha Z')$}\,:\,\pi\models^{L}\varphi_4\})=t+(1-t) =1.
\end{equation}
Thus, by (\ref{eq11}) and Lemma \ref{lemma5.4}, (\ref{eq10}) holds.

The ``only if" part. Assume (\ref{eq10}) holds. By Lemma \ref{lemma5.4}, $$\mathcal{P}(\{\pi\in\text{$RUN(F\alpha Z')$}\,:\,\pi\models^{L}\varphi_3\})+\mathcal{P}(\{\pi\in\text{$RUN(S\alpha Z')$}\,:\,\pi\models^{L}\varphi_4\}) =1. $$

Namely, $\mathcal{P}(F\alpha Z'\models^{L}\varphi_3)=1-\mathcal{P}(S\alpha Z'\models^{L}\varphi_4)=t$ for some $t\in(0,1)\cap\mathbb{Q}$. Together with $\mathcal{P}(N\rightarrow F)=\mathcal{P}(N\rightarrow S)=\frac{1}{2}$ (by (\ref{eq8})) and the fact that the number of configurations reachable from $F\alpha Z'$ (or $S\alpha Z'$) until $Z'$ is at most $2nm$ (which is easy to see), this implies $$\widehat{M_{\triangle}},N\alpha Z' \models^{L}\mathcal{P}_{=\frac{t}{2}}(\varphi_3)\wedge\mathcal{P}_{=\frac{1-t}{2}}(\varphi_4) $$
for any constant $t\in\mathbb{Q}\cap(0,1)$. This completes the proof.
\end{proof}

Now we can prove the following lemma.
\begin{lemma} 
\label{lemma5.6}
For any constant $t\in(0,1)\cap\mathbb{Q}$, $$\widehat{M_{\triangle}},Z\models^{L}\mathcal{P}_{>0}({\bf true}{\bf U}^{\leq 2nm} [C\wedge \mathcal{P}_{=1}({\bf X}[\mathcal{P}_{=\frac{t}{2}}(\varphi_3)\wedge\mathcal{P}_{=\frac{1-t}{2}}(\varphi_4)])] )$$ if and only if $\alpha$ ($|\alpha|\leq nm$) is a solution of the modified bounded PCP instance.
\end{lemma}
\begin{proof}
Let $\pi$ be a path of $\omega$-pBPA $\triangle$ starting from $C$ and induced by $C\alpha Z'$, where $\alpha$ is a guessed candidate solution. Then, 
$$\aligned
(\ref{eq10})&\text{ is true}\\
&\Leftrightarrow\,\, \widehat{M_{\triangle}}, N\alpha Z'\models^{L}\mathcal{P}_{=\frac{t}{2}}(\varphi_3)\wedge\mathcal{P}_{\frac{1-t}{2}}(\varphi_4)\quad\text{( by Lemma \ref{lemma5.5} )}\\
&\Leftrightarrow\,\,\widehat{M_{\triangle}},C\alpha Z\models^{L}{\bf X}[\mathcal{P}_{=\frac{t}{2}}(\varphi_3)\wedge\mathcal{P}_{=\frac{1-t}{2}}(\varphi_4)]\quad\text{( by $C\rightarrow N$ )}\\
&\Leftrightarrow\,\,\widehat{M_{\triangle}},C\models^{L}\mathcal{P}_{=1}({\bf X}[\mathcal{P}_{=\frac{t}{2}}(\varphi_3)\wedge\mathcal{P}_{=\frac{1-t}{2}}(\varphi_4)])\quad\text{( by $\mathcal{P}(C\rightarrow N)=1$)}\\
&\Leftrightarrow\,\,\widehat{M_{\triangle}},Z\models^{L}\mathcal{P}_{>0}({\bf true}{\bf U}^{\leq 2nm}[C\wedge \mathcal{P}_{=1}({\bf X}[\mathcal{P}_{=\frac{t}{2}}(\varphi_3)\wedge\mathcal{P}_{=\frac{1-t}{2}}(\varphi_4)])])\quad\text{( by Lemma \ref{lemma5.1} )}
\endaligned$$

Thus, for any constant $t\in(0,1)\cap\mathbb{Q}$ and any modified bounded PCP instance $\alpha$, the formula
\begin{equation}
\label{eq12}
\begin{split}
\widehat{M_{\triangle}},Z\models^{L}\mathcal{P}_{>0}({\bf true}{\bf U}^{\leq 2nm}[C\wedge \mathcal{P}_{=1}({\bf X}[\mathcal{P}_{=\frac{t}{2}}(\varphi_3)\wedge\mathcal{P}_{=\frac{1-t}{2}}(\varphi_4)])])
\end{split}
\end{equation}
holds if and only if $\alpha$ is a solution of the given modified bounded PCP instance. Consequently, an algorithm deciding whether (\ref{eq12}) holds yields an algorithm for the modified bounded Post Correspondence Problem. Moreover, the above reduction can clearly be performed in deterministic polynomial time.
\end{proof}

Theorem \ref{theorem2} can now be proved as follows:

\subsection{Proof of Theorem \ref{theorem2}}
By Theorem \ref{theorem5}, model checking stateless probabilistic $\omega$-pushdown systems against $\omega$-bounded probabilistic computational tree logic ($\omega$-bPCTL) is decidable. However, it is not known whether this problem lies in $\mathit{NP}$.

By Lemma \ref{lemma5.6} and Theorem \ref{theorem5.2}, Theorem \ref{theorem2} follows.\Q.E.D

\begin{remark}
\label{remark5.2}
Some readers may consider our lower bounds (Theorem \ref{theorem2}) rather weak, given that the qualitative fragment of standard PCTL is already $\mathcal{EXP}$-complete \cite{BBFK14} ($\mathcal{EXP}$ is the set of all decision problems that are solvable by some deterministic Turing machines in exponential time, i.e., in $O(2^{p(n)})$ time, where $p(n)$ is a polynomial function of $n$). However, this comparison is misleading. Because one cannot conclude that model checking $\omega$-pBPA against $\omega$-bPCTL is $\mathcal{EXP}$-hard from the fact that the qualitative fragment of standard PCTL is already $\mathcal{EXP}$-complete \cite{BBFK14}. In fact, we find that the hardness proof in \cite{BBFK14} relies heavily on PCTL formulas that use the ${\bf U}$ operator, which is not permitted in $\omega$-bPCTL. Therefore, their $\mathcal{EXP}$-hardness result does not apply in our setting. In summary, our lower-bound (Theorem \ref{theorem2}) and the $\mathcal{EXP}$-hardness result for the qualitative fragment of standard PCTL \cite{BBFK14} are incomparable in nature. We  can thus conclude that there is currently no known lower bound for this problem that is better than ours.
\end{remark}

\section{Discussion on Upper Bounds for Model Checking $\omega$-pBPA against $\omega$-bPCTL}
\label{discussion_upper_bound}

In this section, we discuss the main difficulties we encountered when attempting to establish an unconditional upper bound for the model-checking problem of $\omega$-pBPA against $\omega$-bPCTL.

Let $\triangle=(\Gamma, \delta,Z,F_{\rm inal},{\bf P}, \mathcal{F})$ be a fixed $\omega$-pBPA, and $\phi$ an $\omega$-bPCTL formula. We denote the input by $(\triangle, \phi)$. The length of the input $(\triangle,\phi)$, denoted by $|(\triangle,\phi)|$, is the number of bits required to encode $\triangle$ and $\phi$ in binary (see e.g.,\cite{DBB12}). We assume this length is $n$, i.e., $n=|(\triangle,\phi)|$.

We first have the following auxiliary result.

\begin{lemma}[cf. Lemma 2.3 in \cite{VW94}]
Under simple assignment, the Büchi accepting condition holds if and only if there is an accepting head in $F_{\rm inal}$ of $\triangle$ that is reachable from $Z$ and reachable from itself.
\end{lemma}

\begin{proof}
The proof is similar to that of Lemma 2.3 in \cite{VW94}. We omit the details.
\end{proof}

\begin{theorem}
\label{theorem9}
The problem of determining whether an accepting head in $F_{\rm inal}$ of $\triangle$ is reachable from itself is in $\mathbf{L}$, where $\mathbf{L}\overset{\rm def}{=}\mathcal{DSPACE}(\log n)$. 
\end{theorem}
\begin{proof}
The proof can be adapted from that of Theorem 2.4 in \cite{VW94} in a manner similar to the deterministic simulation of a nondeterministic Turing machine (see the proof of Theorem 1.9 in \cite{DK14}). Specifically, let $x$ be an accepting head in $F_{\rm inal}$. The following is the algorithm to simulate a computational path.
\begin{enumerate}
  \item [1.]{Make $x$ the current configuration.}
  \item [2.]{Choose a transition from the current configuration and replace the current configuration by the head of the target configuration of this transition. If there is no transition from the current configuration, then return ${\rm NO}$.}
  \item [3.]{If the new configuration is $x$, stop and return ${\rm YES}$. Otherwise repeat from step (2).}
\end{enumerate}

The above algorithm is guaranteed to terminate because there are only finitely many transition rules in any $\omega$-pBPA $\triangle=(\Gamma, \delta,Z,F_{\rm inal},{\bf P}, \mathcal{F})$.

At each step, only three configuration heads (i.e., elements in $\Gamma$) are remembered. Thus, the algorithm requires only deterministic logarithmic space. If the above algorithm return ${\rm YES}$, then we are done; otherwise we proceed to the next simulation. To simulate all computational paths, one needs to erase the space after the previous simulation ends and before the next simulation begins.
\end{proof}
\begin{remark}
\label{remark6.1}
In fact, under a simple assignment $L$, the satisfaction relation of the form $\gamma\models^{L}\mathcal{P}_{\bowtie r}(\varphi^{\omega})$ (where $r\in[0,1]\cap\mathbb{Q}$, $\varphi^{\omega}$ is an infinitary path formula in $\omega$-PCTL, and $\bowtie\in\{>,<,=,\geq,\leq\}$) can also be decided within $\mathbf{L}$, with the proof idea similar to the combination of the proof ideas of Theorem \ref{theorem9} and Theorem \ref{theorem10} (i.e., deterministic simulation of all possible computation paths). 
\end{remark}

To determine the complexity of the model-checking problem, the key objects to study are path formulas of the form $\phi_1{\bf U}^{\leq k}\phi_2$.

\begin{theorem}[Restatement of Theorem \ref{theorem.three}]
\label{theorem10}
For any $\omega$-bPCTL path formula $\phi$, if there exists a function $p(n)$ such that, for any $r=(Z)(\gamma_2)(\gamma_3)\cdots$ in $\triangle$, one can decide whether the prefix $(Z)(\gamma_2)\cdots (\gamma_{p(n)})$ of $r$ satisfies $$(Z)(\gamma_2)\cdots (\gamma_{p(n)})\models^{L}\phi,$$ where $n=|(\triangle,\phi)|$, then the model-checking problem for $\omega$-pBPA against $\omega$-bPCTL is in $\mathcal{DSPACE}(p(n))$.
\end{theorem}
\begin{proof}
We can construct a multi-tape deterministic $p(n)$ space-bounded universal Turing machine $U$ that simulate all runs of $\triangle$ (it is sufficient to simulate only the prefix $(Z)(\gamma_2)\cdots(\gamma_{p(n)})$ to reach a decision), in a manner similar to the deterministic simulation of a nondeterministic Turing machines (see the proof of Theorem 1.9 in \cite{DK14})\footnote{Concretely, the machine $U$ enumerates all possible computation paths of length at most $p(n)$ of $\triangle$ by reusing tape space. When necessary, it invokes an $O(\log n)$ space-bounded Turing machine to decide whether the computation path satisfies a specific Büchi condition (see e.g., Theorem \ref{theorem9} and Remark \ref{remark6.1}). Notice that $\mathbf{L}\subseteq\mathcal{DSPACE}(p(n))$ whenever $p(n)\geq\log n$, and in what follows what we conjecture is that $p(n)$ is a polynomial.}. In the simulation, $U$ uses $O(\log n)$ space to test whether a run $r$ satisfies the specified Büchi or coBüchi conditions (by Theorem \ref{theorem9}), and accumulates the probability of each run $r$ such that $r\models^L\phi$ under the given accepting condition. At the end of the simulation, $U$ sums these probability to decide whether the total probability satisfies the relation $\bowtie$ with the predefined threshold (where $\bowtie\in\{>,<,=,\geq,\leq\}$).
\end{proof}

\begin{remark}
Obviously, if $p(n)$ is a polynomial, then the problem is in $\mathcal{PSPACE}$ by Theorem \ref{theorem10}. If $p(n)=2^{q(n)}$ for some polynomial $q(n)$, then the problem is in $\mathcal{EXPSPACE}$ by Theorem \ref{theorem10}, where $\mathcal{EXPSPACE}$ is the class of all decision problems solvable by a deterministic Turing machine in exponential space, i.e., in $O(2^{q(n)})$ space for some polynomial $q$.
\end{remark}

The main difficulty at present is that we do not know the growth rate of the function $p(n)$. This leads to the following open question.

\begin{question}
We conjecture that $p(n)$ is a polynomial. That is, for any $\omega$-pBPA $\triangle=(\Gamma, \delta,Z,F_{\rm inal},{\bf P}, \mathcal{F})$, any $\omega$-bPCTL path formula $\phi$, and  any run $r=(Z)(\gamma_2)(\gamma_3)\cdots$ in $\triangle$, one can decide whether the prefix $(Z)(\gamma_2)\cdots (\gamma_{p(n)})$ of $r$ satisfies $$(Z)(\gamma_2)\cdots (\gamma_{p(n)})\models^{L}\phi,$$ where $n=|(\triangle,\phi)|$ and $p(n)$ is a polynomial in $n$. 
\end{question}

Determining the exact value (or degree) of $p(n)$ is therefore our primary obstacle.

Note that the state formulas in $\omega$-bPCTL can be highly complex. For example, they may contain nested path formulas such as $\mathcal{P}_{\geq 0.3}(\Psi_1{\bf U}^{\leq k_1}\Psi_2)$, where $\Psi_1$ and $\Psi_2$ themselves contain finitely many nested bounded until $U^{\leq k}$ operators.

\begin{remark}
Can the conjectured $\mathcal{PSPACE}$ upper bound be improved to $\mathcal{NP}$? Such an improvement would be a substantial improvement since $\mathcal{NP}\subseteq\mathcal{PP}\subseteq\mathcal{PSPACE}$, where $\mathcal{PP}$ is the class of decision problems solvable by some probabilistic Turing machines in polynomial time, with an error probability of less than $\frac{1}{2}$ for all instances \cite{DK14}. But this is more difficult than proving a $\mathcal{PSPACE}$ upper bound, and we currently have no answer to this question.
\end{remark}

\section{Concluding Remarks and Open Problems}
\label{sec:conclusion}

To summarize, we first study the expressiveness of the logics PCTL, PCTL$^*$, $\omega$-PCTL, and $\omega$-PCTL$^*$ (Theorem \ref{theorem.seven}). We then define the notion of {\em probabilistic $\omega$-pushdown automata} for the first time in this paper and investigate how Büchi conditions influence a given $\omega$-PCTL formula (Subsection \ref{subsection.three.five}), showing that Büchi conditions determine which runs contribute to the probability calculation in $\omega$-PCTL. At the same time, we also compared the specifications of probabilistic $\omega$-pushdown systems with those expressible by $\omega$-PCTL (Theorem \ref{theorem.seven}).

We then focus mainly on the computational complexity of model checking stateless probabilistic $\omega$-pushdown systems against $\omega$-PCTL. We show that this problem is undecidable for $\omega$-pBPA, which has several corollaries such as Corollary \ref{corollary2} and Corollary \ref{corollary3}.

We next study the complexity of model checking stateless probabilistic $\omega$-pushdown systems against $\omega$-bounded probabilistic computational tree logic ($\omega$-bPCTL) and show that the problem is decidable and $\mathit{NP}$-hard.

We also prove a conditional upper bound for model checking stateless probabilistic $\omega$-pushdown systems against $\omega$-bPCTL and discuss the main difficulties encountered when attempting to solve this problem.

Finally, apart from the open problems listed in Section \ref{discussion_upper_bound} for future research, the satisfiability problem for $\omega$-PCTL remains open. Like the satisfiability problem for LTL, which is known to be $\mathit{PSPACE}$-hard \cite{SC85}(see \cite{BK08}, p. 296), the satisfiability problem for $\omega$-PCTL is defined as follows: given an $\omega$-PCTL state formula $\varphi$, does there exist a probabilistic $\omega$-pushdown system $\triangle$ such that $\widehat{M_{\triangle}},Z\models^L\varphi$? We currently do not know the answer to this intriguing question.

\end{document}